\documentclass[longauth]{aa}

\usepackage{graphicx}
\usepackage{natbib}
\usepackage{scalerel}
\usepackage{ulem}
\usepackage[table]{xcolor}
\usepackage{multirow}
\usepackage{txfonts}
\usepackage[pdfencoding=auto,psdextra]{hyperref}
\hypersetup{
    colorlinks=true,
    linkcolor=blue,
    filecolor=magenta,      
    urlcolor=blue,
    citecolor=blue
}
\usepackage[nameinlink,capitalise]{cleveref}

\makeatletter
\renewcommand*\aa@pageof{, page \thepage{} of \pageref*{LastPage}}
\makeatother

\usepackage[utf8]{inputenc}

\usepackage[switch, modulo]{lineno}

\usepackage{euclid}

\let\Gamma\varGamma
\let\Delta\varDelta
\let\Theta\varTheta
\let\Lambda\varLambda
\let\Xi\varXi
\let\Pi\varPi
\let\Upsilon\varUpsilon
\let\Phi\varPhi
\let\Psi\varPsi
\let\Omega\varOmega

\newcommand{\HH}{{\cal H}}

\newcommand{\al}{\alpha}
\newcommand{\bal}{\vec\alpha}

\newcommand{\de}{\delta}

\newcommand{\ka}{\kappa}

\newcommand{\bean}{\begin{eqnarray}}
\newcommand{\eean}{\end{eqnarray}}
\newcommand{\dd}{\partial}

\newcommand{\ii}{\mathrm{i}}
\newcommand{\ud}{\mathrm{d}}
\newcommand{\udln}{\mathrm{dln}}
\newcommand{\er}{\mathrm{e}}
\newcommand{\id}{{\rm 1\kern -2.5pt I}} 
\newcommand{\real}{\mathbb{R}} 
\newcommand{\bn}{\hat{{\vec n}}}

\newcommand{\bk}{{\vec k}}

\newcommand{\bx}{{\vec x}}
\newcommand{\cd}{{\cdot}}
\newcommand{\bde}{\hat{{\vec e}}}
\newcommand{\bel}{{\vec\ell}}

\begin{document}

\title{\Euclid preparation}
\subtitle{The flat-sky approximation for the clustering of \Euclid's photometric galaxies}  
\newcommand{\orcid}[1]{} 
\author{Euclid Collaboration: W.~L.~Matthewson\orcid{0000-0001-6957-772X}\thanks{\email{willmatt4th@kasi.re.kr}}\inst{\ref{aff1},\ref{aff2}}
\and R.~Durrer\orcid{0000-0001-9833-2086}\inst{\ref{aff1}}
\and S.~Camera\orcid{0000-0003-3399-3574}\inst{\ref{aff3},\ref{aff4},\ref{aff5}}
\and I.~Tutusaus\orcid{0000-0002-3199-0399}\inst{\ref{aff6},\ref{aff7},\ref{aff8}}
\and B.~Altieri\orcid{0000-0003-3936-0284}\inst{\ref{aff9}}
\and A.~Amara\inst{\ref{aff10}}
\and S.~Andreon\orcid{0000-0002-2041-8784}\inst{\ref{aff11}}
\and N.~Auricchio\orcid{0000-0003-4444-8651}\inst{\ref{aff12}}
\and C.~Baccigalupi\orcid{0000-0002-8211-1630}\inst{\ref{aff13},\ref{aff14},\ref{aff15},\ref{aff16}}
\and M.~Baldi\orcid{0000-0003-4145-1943}\inst{\ref{aff17},\ref{aff12},\ref{aff18}}
\and S.~Bardelli\orcid{0000-0002-8900-0298}\inst{\ref{aff12}}
\and P.~Battaglia\orcid{0000-0002-7337-5909}\inst{\ref{aff12}}
\and A.~Biviano\orcid{0000-0002-0857-0732}\inst{\ref{aff14},\ref{aff13}}
\and E.~Branchini\orcid{0000-0002-0808-6908}\inst{\ref{aff19},\ref{aff20},\ref{aff11}}
\and M.~Brescia\orcid{0000-0001-9506-5680}\inst{\ref{aff21},\ref{aff22}}
\and G.~Ca\~nas-Herrera\orcid{0000-0003-2796-2149}\inst{\ref{aff23},\ref{aff24},\ref{aff25}}
\and V.~Capobianco\orcid{0000-0002-3309-7692}\inst{\ref{aff5}}
\and C.~Carbone\orcid{0000-0003-0125-3563}\inst{\ref{aff26}}
\and V.~F.~Cardone\inst{\ref{aff27},\ref{aff28}}
\and J.~Carretero\orcid{0000-0002-3130-0204}\inst{\ref{aff29},\ref{aff30}}
\and S.~Casas\orcid{0000-0002-4751-5138}\inst{\ref{aff31}}
\and M.~Castellano\orcid{0000-0001-9875-8263}\inst{\ref{aff27}}
\and G.~Castignani\orcid{0000-0001-6831-0687}\inst{\ref{aff12}}
\and S.~Cavuoti\orcid{0000-0002-3787-4196}\inst{\ref{aff22},\ref{aff32}}
\and K.~C.~Chambers\orcid{0000-0001-6965-7789}\inst{\ref{aff33}}
\and A.~Cimatti\inst{\ref{aff34}}
\and C.~Colodro-Conde\inst{\ref{aff35}}
\and G.~Congedo\orcid{0000-0003-2508-0046}\inst{\ref{aff36}}
\and C.~J.~Conselice\orcid{0000-0003-1949-7638}\inst{\ref{aff37}}
\and L.~Conversi\orcid{0000-0002-6710-8476}\inst{\ref{aff38},\ref{aff9}}
\and Y.~Copin\orcid{0000-0002-5317-7518}\inst{\ref{aff39}}
\and F.~Courbin\orcid{0000-0003-0758-6510}\inst{\ref{aff40},\ref{aff41}}
\and H.~M.~Courtois\orcid{0000-0003-0509-1776}\inst{\ref{aff42}}
\and A.~Da~Silva\orcid{0000-0002-6385-1609}\inst{\ref{aff43},\ref{aff44}}
\and H.~Degaudenzi\orcid{0000-0002-5887-6799}\inst{\ref{aff45}}
\and G.~De~Lucia\orcid{0000-0002-6220-9104}\inst{\ref{aff14}}
\and H.~Dole\orcid{0000-0002-9767-3839}\inst{\ref{aff46}}
\and F.~Dubath\orcid{0000-0002-6533-2810}\inst{\ref{aff45}}
\and C.~A.~J.~Duncan\orcid{0009-0003-3573-0791}\inst{\ref{aff36},\ref{aff37}}
\and X.~Dupac\inst{\ref{aff9}}
\and S.~Dusini\orcid{0000-0002-1128-0664}\inst{\ref{aff47}}
\and S.~Escoffier\orcid{0000-0002-2847-7498}\inst{\ref{aff48}}
\and M.~Farina\orcid{0000-0002-3089-7846}\inst{\ref{aff49}}
\and F.~Faustini\orcid{0000-0001-6274-5145}\inst{\ref{aff27},\ref{aff50}}
\and S.~Ferriol\inst{\ref{aff39}}
\and F.~Finelli\orcid{0000-0002-6694-3269}\inst{\ref{aff12},\ref{aff51}}
\and M.~Frailis\orcid{0000-0002-7400-2135}\inst{\ref{aff14}}
\and E.~Franceschi\orcid{0000-0002-0585-6591}\inst{\ref{aff12}}
\and M.~Fumana\orcid{0000-0001-6787-5950}\inst{\ref{aff26}}
\and S.~Galeotta\orcid{0000-0002-3748-5115}\inst{\ref{aff14}}
\and K.~George\orcid{0000-0002-1734-8455}\inst{\ref{aff52}}
\and B.~Gillis\orcid{0000-0002-4478-1270}\inst{\ref{aff36}}
\and C.~Giocoli\orcid{0000-0002-9590-7961}\inst{\ref{aff12},\ref{aff18}}
\and J.~Gracia-Carpio\inst{\ref{aff53}}
\and A.~Grazian\orcid{0000-0002-5688-0663}\inst{\ref{aff54}}
\and F.~Grupp\inst{\ref{aff53},\ref{aff55}}
\and S.~V.~H.~Haugan\orcid{0000-0001-9648-7260}\inst{\ref{aff56}}
\and W.~Holmes\inst{\ref{aff57}}
\and F.~Hormuth\inst{\ref{aff58}}
\and A.~Hornstrup\orcid{0000-0002-3363-0936}\inst{\ref{aff59},\ref{aff60}}
\and K.~Jahnke\orcid{0000-0003-3804-2137}\inst{\ref{aff61}}
\and M.~Jhabvala\inst{\ref{aff62}}
\and B.~Joachimi\orcid{0000-0001-7494-1303}\inst{\ref{aff63}}
\and E.~Keih\"anen\orcid{0000-0003-1804-7715}\inst{\ref{aff64}}
\and S.~Kermiche\orcid{0000-0002-0302-5735}\inst{\ref{aff48}}
\and A.~Kiessling\orcid{0000-0002-2590-1273}\inst{\ref{aff57}}
\and B.~Kubik\orcid{0009-0006-5823-4880}\inst{\ref{aff39}}
\and M.~Kunz\orcid{0000-0002-3052-7394}\inst{\ref{aff1}}
\and H.~Kurki-Suonio\orcid{0000-0002-4618-3063}\inst{\ref{aff65},\ref{aff66}}
\and A.~M.~C.~Le~Brun\orcid{0000-0002-0936-4594}\inst{\ref{aff67}}
\and S.~Ligori\orcid{0000-0003-4172-4606}\inst{\ref{aff5}}
\and P.~B.~Lilje\orcid{0000-0003-4324-7794}\inst{\ref{aff56}}
\and V.~Lindholm\orcid{0000-0003-2317-5471}\inst{\ref{aff65},\ref{aff66}}
\and I.~Lloro\orcid{0000-0001-5966-1434}\inst{\ref{aff68}}
\and G.~Mainetti\orcid{0000-0003-2384-2377}\inst{\ref{aff69}}
\and D.~Maino\inst{\ref{aff70},\ref{aff26},\ref{aff71}}
\and E.~Maiorano\orcid{0000-0003-2593-4355}\inst{\ref{aff12}}
\and O.~Mansutti\orcid{0000-0001-5758-4658}\inst{\ref{aff14}}
\and S.~Marcin\inst{\ref{aff72}}
\and O.~Marggraf\orcid{0000-0001-7242-3852}\inst{\ref{aff73}}
\and M.~Martinelli\orcid{0000-0002-6943-7732}\inst{\ref{aff27},\ref{aff28}}
\and N.~Martinet\orcid{0000-0003-2786-7790}\inst{\ref{aff74}}
\and F.~Marulli\orcid{0000-0002-8850-0303}\inst{\ref{aff75},\ref{aff12},\ref{aff18}}
\and R.~J.~Massey\orcid{0000-0002-6085-3780}\inst{\ref{aff76}}
\and E.~Medinaceli\orcid{0000-0002-4040-7783}\inst{\ref{aff12}}
\and S.~Mei\orcid{0000-0002-2849-559X}\inst{\ref{aff77},\ref{aff78}}
\and Y.~Mellier\inst{\ref{aff79},\ref{aff80}}
\and M.~Meneghetti\orcid{0000-0003-1225-7084}\inst{\ref{aff12},\ref{aff18}}
\and E.~Merlin\orcid{0000-0001-6870-8900}\inst{\ref{aff27}}
\and G.~Meylan\inst{\ref{aff81}}
\and A.~Mora\orcid{0000-0002-1922-8529}\inst{\ref{aff82}}
\and M.~Moresco\orcid{0000-0002-7616-7136}\inst{\ref{aff75},\ref{aff12}}
\and B.~Morin\inst{\ref{aff83}}
\and L.~Moscardini\orcid{0000-0002-3473-6716}\inst{\ref{aff75},\ref{aff12},\ref{aff18}}
\and C.~Neissner\orcid{0000-0001-8524-4968}\inst{\ref{aff84},\ref{aff30}}
\and S.-M.~Niemi\orcid{0009-0005-0247-0086}\inst{\ref{aff23}}
\and C.~Padilla\orcid{0000-0001-7951-0166}\inst{\ref{aff84}}
\and S.~Paltani\orcid{0000-0002-8108-9179}\inst{\ref{aff45}}
\and F.~Pasian\orcid{0000-0002-4869-3227}\inst{\ref{aff14}}
\and K.~Pedersen\inst{\ref{aff85}}
\and W.~J.~Percival\orcid{0000-0002-0644-5727}\inst{\ref{aff86},\ref{aff87},\ref{aff88}}
\and V.~Pettorino\orcid{0000-0002-4203-9320}\inst{\ref{aff23}}
\and S.~Pires\orcid{0000-0002-0249-2104}\inst{\ref{aff83}}
\and G.~Polenta\orcid{0000-0003-4067-9196}\inst{\ref{aff50}}
\and M.~Poncet\inst{\ref{aff89}}
\and L.~A.~Popa\inst{\ref{aff90}}
\and F.~Raison\orcid{0000-0002-7819-6918}\inst{\ref{aff53}}
\and R.~Rebolo\orcid{0000-0003-3767-7085}\inst{\ref{aff35},\ref{aff91},\ref{aff92}}
\and A.~Renzi\orcid{0000-0001-9856-1970}\inst{\ref{aff93},\ref{aff47}}
\and J.~Rhodes\orcid{0000-0002-4485-8549}\inst{\ref{aff57}}
\and G.~Riccio\inst{\ref{aff22}}
\and E.~Romelli\orcid{0000-0003-3069-9222}\inst{\ref{aff14}}
\and M.~Roncarelli\orcid{0000-0001-9587-7822}\inst{\ref{aff12}}
\and R.~Saglia\orcid{0000-0003-0378-7032}\inst{\ref{aff55},\ref{aff53}}
\and Z.~Sakr\orcid{0000-0002-4823-3757}\inst{\ref{aff94},\ref{aff8},\ref{aff95}}
\and A.~G.~S\'anchez\orcid{0000-0003-1198-831X}\inst{\ref{aff53}}
\and D.~Sapone\orcid{0000-0001-7089-4503}\inst{\ref{aff96}}
\and B.~Sartoris\orcid{0000-0003-1337-5269}\inst{\ref{aff55},\ref{aff14}}
\and P.~Schneider\orcid{0000-0001-8561-2679}\inst{\ref{aff73}}
\and T.~Schrabback\orcid{0000-0002-6987-7834}\inst{\ref{aff97}}
\and A.~Secroun\orcid{0000-0003-0505-3710}\inst{\ref{aff48}}
\and E.~Sefusatti\orcid{0000-0003-0473-1567}\inst{\ref{aff14},\ref{aff13},\ref{aff15}}
\and G.~Seidel\orcid{0000-0003-2907-353X}\inst{\ref{aff61}}
\and S.~Serrano\orcid{0000-0002-0211-2861}\inst{\ref{aff7},\ref{aff98},\ref{aff6}}
\and P.~Simon\inst{\ref{aff73}}
\and C.~Sirignano\orcid{0000-0002-0995-7146}\inst{\ref{aff93},\ref{aff47}}
\and G.~Sirri\orcid{0000-0003-2626-2853}\inst{\ref{aff18}}
\and A.~Spurio~Mancini\orcid{0000-0001-5698-0990}\inst{\ref{aff99}}
\and L.~Stanco\orcid{0000-0002-9706-5104}\inst{\ref{aff47}}
\and J.-L.~Starck\orcid{0000-0003-2177-7794}\inst{\ref{aff83}}
\and J.~Steinwagner\orcid{0000-0001-7443-1047}\inst{\ref{aff53}}
\and P.~Tallada-Cresp\'{i}\orcid{0000-0002-1336-8328}\inst{\ref{aff29},\ref{aff30}}
\and A.~N.~Taylor\inst{\ref{aff36}}
\and I.~Tereno\orcid{0000-0002-4537-6218}\inst{\ref{aff43},\ref{aff100}}
\and N.~Tessore\orcid{0000-0002-9696-7931}\inst{\ref{aff63}}
\and S.~Toft\orcid{0000-0003-3631-7176}\inst{\ref{aff101},\ref{aff102}}
\and R.~Toledo-Moreo\orcid{0000-0002-2997-4859}\inst{\ref{aff103}}
\and F.~Torradeflot\orcid{0000-0003-1160-1517}\inst{\ref{aff30},\ref{aff29}}
\and L.~Valenziano\orcid{0000-0002-1170-0104}\inst{\ref{aff12},\ref{aff51}}
\and J.~Valiviita\orcid{0000-0001-6225-3693}\inst{\ref{aff65},\ref{aff66}}
\and T.~Vassallo\orcid{0000-0001-6512-6358}\inst{\ref{aff14},\ref{aff52}}
\and A.~Veropalumbo\orcid{0000-0003-2387-1194}\inst{\ref{aff11},\ref{aff20},\ref{aff19}}
\and Y.~Wang\orcid{0000-0002-4749-2984}\inst{\ref{aff104}}
\and J.~Weller\orcid{0000-0002-8282-2010}\inst{\ref{aff55},\ref{aff53}}
\and G.~Zamorani\orcid{0000-0002-2318-301X}\inst{\ref{aff12}}
\and E.~Zucca\orcid{0000-0002-5845-8132}\inst{\ref{aff12}}
\and M.~Ballardini\orcid{0000-0003-4481-3559}\inst{\ref{aff105},\ref{aff106},\ref{aff12}}
\and E.~Bozzo\orcid{0000-0002-8201-1525}\inst{\ref{aff45}}
\and C.~Burigana\orcid{0000-0002-3005-5796}\inst{\ref{aff107},\ref{aff51}}
\and R.~Cabanac\orcid{0000-0001-6679-2600}\inst{\ref{aff8}}
\and M.~Calabrese\orcid{0000-0002-2637-2422}\inst{\ref{aff108},\ref{aff26}}
\and A.~Cappi\inst{\ref{aff12},\ref{aff109}}
\and D.~Di~Ferdinando\inst{\ref{aff18}}
\and J.~A.~Escartin~Vigo\inst{\ref{aff53}}
\and L.~Gabarra\orcid{0000-0002-8486-8856}\inst{\ref{aff110}}
\and W.~G.~Hartley\inst{\ref{aff45}}
\and J.~Mart\'{i}n-Fleitas\orcid{0000-0002-8594-569X}\inst{\ref{aff111}}
\and S.~Matthew\orcid{0000-0001-8448-1697}\inst{\ref{aff36}}
\and M.~Maturi\orcid{0000-0002-3517-2422}\inst{\ref{aff94},\ref{aff112}}
\and N.~Mauri\orcid{0000-0001-8196-1548}\inst{\ref{aff34},\ref{aff18}}
\and R.~B.~Metcalf\orcid{0000-0003-3167-2574}\inst{\ref{aff75},\ref{aff12}}
\and A.~Pezzotta\orcid{0000-0003-0726-2268}\inst{\ref{aff11}}
\and M.~P\"ontinen\orcid{0000-0001-5442-2530}\inst{\ref{aff65}}
\and C.~Porciani\orcid{0000-0002-7797-2508}\inst{\ref{aff73}}
\and I.~Risso\orcid{0000-0003-2525-7761}\inst{\ref{aff11},\ref{aff20}}
\and V.~Scottez\orcid{0009-0008-3864-940X}\inst{\ref{aff79},\ref{aff113}}
\and M.~Sereno\orcid{0000-0003-0302-0325}\inst{\ref{aff12},\ref{aff18}}
\and M.~Tenti\orcid{0000-0002-4254-5901}\inst{\ref{aff18}}
\and M.~Viel\orcid{0000-0002-2642-5707}\inst{\ref{aff13},\ref{aff14},\ref{aff16},\ref{aff15},\ref{aff114}}
\and M.~Wiesmann\orcid{0009-0000-8199-5860}\inst{\ref{aff56}}
\and Y.~Akrami\orcid{0000-0002-2407-7956}\inst{\ref{aff115},\ref{aff116}}
\and S.~Alvi\orcid{0000-0001-5779-8568}\inst{\ref{aff105}}
\and I.~T.~Andika\orcid{0000-0001-6102-9526}\inst{\ref{aff117},\ref{aff118}}
\and S.~Anselmi\orcid{0000-0002-3579-9583}\inst{\ref{aff47},\ref{aff93},\ref{aff119}}
\and M.~Archidiacono\orcid{0000-0003-4952-9012}\inst{\ref{aff70},\ref{aff71}}
\and F.~Atrio-Barandela\orcid{0000-0002-2130-2513}\inst{\ref{aff120}}
\and D.~Bertacca\orcid{0000-0002-2490-7139}\inst{\ref{aff93},\ref{aff54},\ref{aff47}}
\and M.~Bethermin\orcid{0000-0002-3915-2015}\inst{\ref{aff121}}
\and L.~Blot\orcid{0000-0002-9622-7167}\inst{\ref{aff122},\ref{aff67}}
\and M.~Bonici\orcid{0000-0002-8430-126X}\inst{\ref{aff86},\ref{aff26}}
\and S.~Borgani\orcid{0000-0001-6151-6439}\inst{\ref{aff123},\ref{aff13},\ref{aff14},\ref{aff15},\ref{aff114}}
\and M.~L.~Brown\orcid{0000-0002-0370-8077}\inst{\ref{aff37}}
\and S.~Bruton\orcid{0000-0002-6503-5218}\inst{\ref{aff124}}
\and A.~Calabro\orcid{0000-0003-2536-1614}\inst{\ref{aff27}}
\and B.~Camacho~Quevedo\orcid{0000-0002-8789-4232}\inst{\ref{aff13},\ref{aff16},\ref{aff14}}
\and F.~Caro\inst{\ref{aff27}}
\and C.~S.~Carvalho\inst{\ref{aff100}}
\and T.~Castro\orcid{0000-0002-6292-3228}\inst{\ref{aff14},\ref{aff15},\ref{aff13},\ref{aff114}}
\and F.~Cogato\orcid{0000-0003-4632-6113}\inst{\ref{aff75},\ref{aff12}}
\and S.~Conseil\orcid{0000-0002-3657-4191}\inst{\ref{aff39}}
\and A.~R.~Cooray\orcid{0000-0002-3892-0190}\inst{\ref{aff125}}
\and S.~Davini\orcid{0000-0003-3269-1718}\inst{\ref{aff20}}
\and G.~Desprez\orcid{0000-0001-8325-1742}\inst{\ref{aff126}}
\and A.~D\'iaz-S\'anchez\orcid{0000-0003-0748-4768}\inst{\ref{aff127}}
\and J.~J.~Diaz\orcid{0000-0003-2101-1078}\inst{\ref{aff35}}
\and S.~Di~Domizio\orcid{0000-0003-2863-5895}\inst{\ref{aff19},\ref{aff20}}
\and J.~M.~Diego\orcid{0000-0001-9065-3926}\inst{\ref{aff128}}
\and M.~Y.~Elkhashab\orcid{0000-0001-9306-2603}\inst{\ref{aff14},\ref{aff15},\ref{aff123},\ref{aff13}}
\and A.~Enia\orcid{0000-0002-0200-2857}\inst{\ref{aff17},\ref{aff12}}
\and Y.~Fang\orcid{0000-0002-0334-6950}\inst{\ref{aff55}}
\and A.~G.~Ferrari\orcid{0009-0005-5266-4110}\inst{\ref{aff18}}
\and A.~Finoguenov\orcid{0000-0002-4606-5403}\inst{\ref{aff65}}
\and A.~Franco\orcid{0000-0002-4761-366X}\inst{\ref{aff129},\ref{aff130},\ref{aff131}}
\and K.~Ganga\orcid{0000-0001-8159-8208}\inst{\ref{aff77}}
\and J.~Garc\'ia-Bellido\orcid{0000-0002-9370-8360}\inst{\ref{aff115}}
\and T.~Gasparetto\orcid{0000-0002-7913-4866}\inst{\ref{aff27}}
\and V.~Gautard\inst{\ref{aff132}}
\and E.~Gaztanaga\orcid{0000-0001-9632-0815}\inst{\ref{aff6},\ref{aff7},\ref{aff133}}
\and F.~Giacomini\orcid{0000-0002-3129-2814}\inst{\ref{aff18}}
\and F.~Gianotti\orcid{0000-0003-4666-119X}\inst{\ref{aff12}}
\and G.~Gozaliasl\orcid{0000-0002-0236-919X}\inst{\ref{aff134},\ref{aff65}}
\and C.~M.~Gutierrez\orcid{0000-0001-7854-783X}\inst{\ref{aff135}}
\and S.~Hemmati\orcid{0000-0003-2226-5395}\inst{\ref{aff136}}
\and C.~Hern\'andez-Monteagudo\orcid{0000-0001-5471-9166}\inst{\ref{aff92},\ref{aff35}}
\and H.~Hildebrandt\orcid{0000-0002-9814-3338}\inst{\ref{aff137}}
\and J.~Hjorth\orcid{0000-0002-4571-2306}\inst{\ref{aff85}}
\and J.~J.~E.~Kajava\orcid{0000-0002-3010-8333}\inst{\ref{aff138},\ref{aff139}}
\and Y.~Kang\orcid{0009-0000-8588-7250}\inst{\ref{aff45}}
\and V.~Kansal\orcid{0000-0002-4008-6078}\inst{\ref{aff140},\ref{aff141}}
\and D.~Karagiannis\orcid{0000-0002-4927-0816}\inst{\ref{aff105},\ref{aff142}}
\and K.~Kiiveri\inst{\ref{aff64}}
\and J.~Kim\orcid{0000-0003-2776-2761}\inst{\ref{aff110}}
\and C.~C.~Kirkpatrick\inst{\ref{aff64}}
\and S.~Kruk\orcid{0000-0001-8010-8879}\inst{\ref{aff9}}
\and F.~Lacasa\orcid{0000-0002-7268-3440}\inst{\ref{aff143},\ref{aff46}}
\and M.~Lattanzi\orcid{0000-0003-1059-2532}\inst{\ref{aff106}}
\and J.~Le~Graet\orcid{0000-0001-6523-7971}\inst{\ref{aff48}}
\and L.~Legrand\orcid{0000-0003-0610-5252}\inst{\ref{aff144},\ref{aff145}}
\and M.~Lembo\orcid{0000-0002-5271-5070}\inst{\ref{aff80}}
\and F.~Lepori\orcid{0009-0000-5061-7138}\inst{\ref{aff146}}
\and G.~Leroy\orcid{0009-0004-2523-4425}\inst{\ref{aff147},\ref{aff76}}
\and G.~F.~Lesci\orcid{0000-0002-4607-2830}\inst{\ref{aff75},\ref{aff12}}
\and J.~Lesgourgues\orcid{0000-0001-7627-353X}\inst{\ref{aff31}}
\and T.~I.~Liaudat\orcid{0000-0002-9104-314X}\inst{\ref{aff148}}
\and J.~Macias-Perez\orcid{0000-0002-5385-2763}\inst{\ref{aff149}}
\and G.~Maggio\orcid{0000-0003-4020-4836}\inst{\ref{aff14}}
\and M.~Magliocchetti\orcid{0000-0001-9158-4838}\inst{\ref{aff49}}
\and R.~Maoli\orcid{0000-0002-6065-3025}\inst{\ref{aff150},\ref{aff27}}
\and C.~J.~A.~P.~Martins\orcid{0000-0002-4886-9261}\inst{\ref{aff151},\ref{aff152}}
\and L.~Maurin\orcid{0000-0002-8406-0857}\inst{\ref{aff46}}
\and M.~Miluzio\inst{\ref{aff9},\ref{aff153}}
\and P.~Monaco\orcid{0000-0003-2083-7564}\inst{\ref{aff123},\ref{aff14},\ref{aff15},\ref{aff13},\ref{aff114}}
\and C.~Moretti\orcid{0000-0003-3314-8936}\inst{\ref{aff14},\ref{aff13},\ref{aff15},\ref{aff16}}
\and G.~Morgante\inst{\ref{aff12}}
\and S.~Nadathur\orcid{0000-0001-9070-3102}\inst{\ref{aff133}}
\and K.~Naidoo\orcid{0000-0002-9182-1802}\inst{\ref{aff133},\ref{aff63}}
\and A.~Navarro-Alsina\orcid{0000-0002-3173-2592}\inst{\ref{aff73}}
\and S.~Nesseris\orcid{0000-0002-0567-0324}\inst{\ref{aff115}}
\and D.~Paoletti\orcid{0000-0003-4761-6147}\inst{\ref{aff12},\ref{aff51}}
\and F.~Passalacqua\orcid{0000-0002-8606-4093}\inst{\ref{aff93},\ref{aff47}}
\and K.~Paterson\orcid{0000-0001-8340-3486}\inst{\ref{aff61}}
\and L.~Patrizii\inst{\ref{aff18}}
\and A.~Pisani\orcid{0000-0002-6146-4437}\inst{\ref{aff48}}
\and D.~Potter\orcid{0000-0002-0757-5195}\inst{\ref{aff146}}
\and S.~Quai\orcid{0000-0002-0449-8163}\inst{\ref{aff75},\ref{aff12}}
\and M.~Radovich\orcid{0000-0002-3585-866X}\inst{\ref{aff54}}
\and G.~Rodighiero\orcid{0000-0002-9415-2296}\inst{\ref{aff93},\ref{aff54}}
\and S.~Sacquegna\orcid{0000-0002-8433-6630}\inst{\ref{aff154},\ref{aff130},\ref{aff129}}
\and M.~Sahl\'en\orcid{0000-0003-0973-4804}\inst{\ref{aff155}}
\and D.~B.~Sanders\orcid{0000-0002-1233-9998}\inst{\ref{aff33}}
\and E.~Sarpa\orcid{0000-0002-1256-655X}\inst{\ref{aff16},\ref{aff114},\ref{aff15}}
\and A.~Schneider\orcid{0000-0001-7055-8104}\inst{\ref{aff146}}
\and D.~Sciotti\orcid{0009-0008-4519-2620}\inst{\ref{aff27},\ref{aff28}}
\and E.~Sellentin\inst{\ref{aff156},\ref{aff25}}
\and A.~Silvestri\orcid{0000-0001-6904-5061}\inst{\ref{aff24}}
\and L.~C.~Smith\orcid{0000-0002-3259-2771}\inst{\ref{aff157}}
\and K.~Tanidis\orcid{0000-0001-9843-5130}\inst{\ref{aff110}}
\and C.~Tao\orcid{0000-0001-7961-8177}\inst{\ref{aff48}}
\and G.~Testera\inst{\ref{aff20}}
\and R.~Teyssier\orcid{0000-0001-7689-0933}\inst{\ref{aff158}}
\and S.~Tosi\orcid{0000-0002-7275-9193}\inst{\ref{aff19},\ref{aff20},\ref{aff11}}
\and A.~Troja\orcid{0000-0003-0239-4595}\inst{\ref{aff93},\ref{aff47}}
\and M.~Tucci\inst{\ref{aff45}}
\and C.~Valieri\inst{\ref{aff18}}
\and A.~Venhola\orcid{0000-0001-6071-4564}\inst{\ref{aff159}}
\and D.~Vergani\orcid{0000-0003-0898-2216}\inst{\ref{aff12}}
\and F.~Vernizzi\orcid{0000-0003-3426-2802}\inst{\ref{aff160}}
\and G.~Verza\orcid{0000-0002-1886-8348}\inst{\ref{aff161}}
\and N.~A.~Walton\orcid{0000-0003-3983-8778}\inst{\ref{aff157}}}
										   
\institute{Universit\'e de Gen\`eve, D\'epartement de Physique Th\'eorique and Centre for Astroparticle Physics, 24 quai Ernest-Ansermet, CH-1211 Gen\`eve 4, Switzerland\label{aff1}
\and
Korea Astronomy and Space Science Institute, 776 Daedeok-daero, Yuseong-gu, Daejeon 34055, Republic of Korea\label{aff2}
\and
Dipartimento di Fisica, Universit\`a degli Studi di Torino, Via P. Giuria 1, 10125 Torino, Italy\label{aff3}
\and
INFN-Sezione di Torino, Via P. Giuria 1, 10125 Torino, Italy\label{aff4}
\and
INAF-Osservatorio Astrofisico di Torino, Via Osservatorio 20, 10025 Pino Torinese (TO), Italy\label{aff5}
\and
Institute of Space Sciences (ICE, CSIC), Campus UAB, Carrer de Can Magrans, s/n, 08193 Barcelona, Spain\label{aff6}
\and
Institut d'Estudis Espacials de Catalunya (IEEC),  Edifici RDIT, Campus UPC, 08860 Castelldefels, Barcelona, Spain\label{aff7}
\and
Institut de Recherche en Astrophysique et Plan\'etologie (IRAP), Universit\'e de Toulouse, CNRS, UPS, CNES, 14 Av. Edouard Belin, 31400 Toulouse, France\label{aff8}
\and
ESAC/ESA, Camino Bajo del Castillo, s/n., Urb. Villafranca del Castillo, 28692 Villanueva de la Ca\~nada, Madrid, Spain\label{aff9}
\and
School of Mathematics and Physics, University of Surrey, Guildford, Surrey, GU2 7XH, UK\label{aff10}
\and
INAF-Osservatorio Astronomico di Brera, Via Brera 28, 20122 Milano, Italy\label{aff11}
\and
INAF-Osservatorio di Astrofisica e Scienza dello Spazio di Bologna, Via Piero Gobetti 93/3, 40129 Bologna, Italy\label{aff12}
\and
IFPU, Institute for Fundamental Physics of the Universe, via Beirut 2, 34151 Trieste, Italy\label{aff13}
\and
INAF-Osservatorio Astronomico di Trieste, Via G. B. Tiepolo 11, 34143 Trieste, Italy\label{aff14}
\and
INFN, Sezione di Trieste, Via Valerio 2, 34127 Trieste TS, Italy\label{aff15}
\and
SISSA, International School for Advanced Studies, Via Bonomea 265, 34136 Trieste TS, Italy\label{aff16}
\and
Dipartimento di Fisica e Astronomia, Universit\`a di Bologna, Via Gobetti 93/2, 40129 Bologna, Italy\label{aff17}
\and
INFN-Sezione di Bologna, Viale Berti Pichat 6/2, 40127 Bologna, Italy\label{aff18}
\and
Dipartimento di Fisica, Universit\`a di Genova, Via Dodecaneso 33, 16146, Genova, Italy\label{aff19}
\and
INFN-Sezione di Genova, Via Dodecaneso 33, 16146, Genova, Italy\label{aff20}
\and
Department of Physics "E. Pancini", University Federico II, Via Cinthia 6, 80126, Napoli, Italy\label{aff21}
\and
INAF-Osservatorio Astronomico di Capodimonte, Via Moiariello 16, 80131 Napoli, Italy\label{aff22}
\and
European Space Agency/ESTEC, Keplerlaan 1, 2201 AZ Noordwijk, The Netherlands\label{aff23}
\and
Institute Lorentz, Leiden University, Niels Bohrweg 2, 2333 CA Leiden, The Netherlands\label{aff24}
\and
Leiden Observatory, Leiden University, Einsteinweg 55, 2333 CC Leiden, The Netherlands\label{aff25}
\and
INAF-IASF Milano, Via Alfonso Corti 12, 20133 Milano, Italy\label{aff26}
\and
INAF-Osservatorio Astronomico di Roma, Via Frascati 33, 00078 Monteporzio Catone, Italy\label{aff27}
\and
INFN-Sezione di Roma, Piazzale Aldo Moro, 2 - c/o Dipartimento di Fisica, Edificio G. Marconi, 00185 Roma, Italy\label{aff28}
\and
Centro de Investigaciones Energ\'eticas, Medioambientales y Tecnol\'ogicas (CIEMAT), Avenida Complutense 40, 28040 Madrid, Spain\label{aff29}
\and
Port d'Informaci\'{o} Cient\'{i}fica, Campus UAB, C. Albareda s/n, 08193 Bellaterra (Barcelona), Spain\label{aff30}
\and
Institute for Theoretical Particle Physics and Cosmology (TTK), RWTH Aachen University, 52056 Aachen, Germany\label{aff31}
\and
INFN section of Naples, Via Cinthia 6, 80126, Napoli, Italy\label{aff32}
\and
Institute for Astronomy, University of Hawaii, 2680 Woodlawn Drive, Honolulu, HI 96822, USA\label{aff33}
\and
Dipartimento di Fisica e Astronomia "Augusto Righi" - Alma Mater Studiorum Universit\`a di Bologna, Viale Berti Pichat 6/2, 40127 Bologna, Italy\label{aff34}
\and
Instituto de Astrof\'{\i}sica de Canarias, E-38205 La Laguna, Tenerife, Spain\label{aff35}
\and
Institute for Astronomy, University of Edinburgh, Royal Observatory, Blackford Hill, Edinburgh EH9 3HJ, UK\label{aff36}
\and
Jodrell Bank Centre for Astrophysics, Department of Physics and Astronomy, University of Manchester, Oxford Road, Manchester M13 9PL, UK\label{aff37}
\and
European Space Agency/ESRIN, Largo Galileo Galilei 1, 00044 Frascati, Roma, Italy\label{aff38}
\and
Universit\'e Claude Bernard Lyon 1, CNRS/IN2P3, IP2I Lyon, UMR 5822, Villeurbanne, F-69100, France\label{aff39}
\and
Institut de Ci\`{e}ncies del Cosmos (ICCUB), Universitat de Barcelona (IEEC-UB), Mart\'{i} i Franqu\`{e}s 1, 08028 Barcelona, Spain\label{aff40}
\and
Instituci\'o Catalana de Recerca i Estudis Avan\c{c}ats (ICREA), Passeig de Llu\'{\i}s Companys 23, 08010 Barcelona, Spain\label{aff41}
\and
UCB Lyon 1, CNRS/IN2P3, IUF, IP2I Lyon, 4 rue Enrico Fermi, 69622 Villeurbanne, France\label{aff42}
\and
Departamento de F\'isica, Faculdade de Ci\^encias, Universidade de Lisboa, Edif\'icio C8, Campo Grande, PT1749-016 Lisboa, Portugal\label{aff43}
\and
Instituto de Astrof\'isica e Ci\^encias do Espa\c{c}o, Faculdade de Ci\^encias, Universidade de Lisboa, Campo Grande, 1749-016 Lisboa, Portugal\label{aff44}
\and
Department of Astronomy, University of Geneva, ch. d'Ecogia 16, 1290 Versoix, Switzerland\label{aff45}
\and
Universit\'e Paris-Saclay, CNRS, Institut d'astrophysique spatiale, 91405, Orsay, France\label{aff46}
\and
INFN-Padova, Via Marzolo 8, 35131 Padova, Italy\label{aff47}
\and
Aix-Marseille Universit\'e, CNRS/IN2P3, CPPM, Marseille, France\label{aff48}
\and
INAF-Istituto di Astrofisica e Planetologia Spaziali, via del Fosso del Cavaliere, 100, 00100 Roma, Italy\label{aff49}
\and
Space Science Data Center, Italian Space Agency, via del Politecnico snc, 00133 Roma, Italy\label{aff50}
\and
INFN-Bologna, Via Irnerio 46, 40126 Bologna, Italy\label{aff51}
\and
University Observatory, LMU Faculty of Physics, Scheinerstrasse 1, 81679 Munich, Germany\label{aff52}
\and
Max Planck Institute for Extraterrestrial Physics, Giessenbachstr. 1, 85748 Garching, Germany\label{aff53}
\and
INAF-Osservatorio Astronomico di Padova, Via dell'Osservatorio 5, 35122 Padova, Italy\label{aff54}
\and
Universit\"ats-Sternwarte M\"unchen, Fakult\"at f\"ur Physik, Ludwig-Maximilians-Universit\"at M\"unchen, Scheinerstrasse 1, 81679 M\"unchen, Germany\label{aff55}
\and
Institute of Theoretical Astrophysics, University of Oslo, P.O. Box 1029 Blindern, 0315 Oslo, Norway\label{aff56}
\and
Jet Propulsion Laboratory, California Institute of Technology, 4800 Oak Grove Drive, Pasadena, CA, 91109, USA\label{aff57}
\and
Felix Hormuth Engineering, Goethestr. 17, 69181 Leimen, Germany\label{aff58}
\and
Technical University of Denmark, Elektrovej 327, 2800 Kgs. Lyngby, Denmark\label{aff59}
\and
Cosmic Dawn Center (DAWN), Denmark\label{aff60}
\and
Max-Planck-Institut f\"ur Astronomie, K\"onigstuhl 17, 69117 Heidelberg, Germany\label{aff61}
\and
NASA Goddard Space Flight Center, Greenbelt, MD 20771, USA\label{aff62}
\and
Department of Physics and Astronomy, University College London, Gower Street, London WC1E 6BT, UK\label{aff63}
\and
Department of Physics and Helsinki Institute of Physics, Gustaf H\"allstr\"omin katu 2, University of Helsinki, 00014 Helsinki, Finland\label{aff64}
\and
Department of Physics, P.O. Box 64, University of Helsinki, 00014 Helsinki, Finland\label{aff65}
\and
Helsinki Institute of Physics, Gustaf H{\"a}llstr{\"o}min katu 2, University of Helsinki, 00014 Helsinki, Finland\label{aff66}
\and
Laboratoire d'etude de l'Univers et des phenomenes eXtremes, Observatoire de Paris, Universit\'e PSL, Sorbonne Universit\'e, CNRS, 92190 Meudon, France\label{aff67}
\and
SKAO, Jodrell Bank, Lower Withington, Macclesfield SK11 9FT, UK\label{aff68}
\and
Centre de Calcul de l'IN2P3/CNRS, 21 avenue Pierre de Coubertin 69627 Villeurbanne Cedex, France\label{aff69}
\and
Dipartimento di Fisica "Aldo Pontremoli", Universit\`a degli Studi di Milano, Via Celoria 16, 20133 Milano, Italy\label{aff70}
\and
INFN-Sezione di Milano, Via Celoria 16, 20133 Milano, Italy\label{aff71}
\and
University of Applied Sciences and Arts of Northwestern Switzerland, School of Computer Science, 5210 Windisch, Switzerland\label{aff72}
\and
Universit\"at Bonn, Argelander-Institut f\"ur Astronomie, Auf dem H\"ugel 71, 53121 Bonn, Germany\label{aff73}
\and
Aix-Marseille Universit\'e, CNRS, CNES, LAM, Marseille, France\label{aff74}
\and
Dipartimento di Fisica e Astronomia "Augusto Righi" - Alma Mater Studiorum Universit\`a di Bologna, via Piero Gobetti 93/2, 40129 Bologna, Italy\label{aff75}
\and
Department of Physics, Institute for Computational Cosmology, Durham University, South Road, Durham, DH1 3LE, UK\label{aff76}
\and
Universit\'e Paris Cit\'e, CNRS, Astroparticule et Cosmologie, 75013 Paris, France\label{aff77}
\and
CNRS-UCB International Research Laboratory, Centre Pierre Bin\'etruy, IRL2007, CPB-IN2P3, Berkeley, USA\label{aff78}
\and
Institut d'Astrophysique de Paris, 98bis Boulevard Arago, 75014, Paris, France\label{aff79}
\and
Institut d'Astrophysique de Paris, UMR 7095, CNRS, and Sorbonne Universit\'e, 98 bis boulevard Arago, 75014 Paris, France\label{aff80}
\and
Institute of Physics, Laboratory of Astrophysics, Ecole Polytechnique F\'ed\'erale de Lausanne (EPFL), Observatoire de Sauverny, 1290 Versoix, Switzerland\label{aff81}
\and
Telespazio UK S.L. for European Space Agency (ESA), Camino bajo del Castillo, s/n, Urbanizacion Villafranca del Castillo, Villanueva de la Ca\~nada, 28692 Madrid, Spain\label{aff82}
\and
Universit\'e Paris-Saclay, Universit\'e Paris Cit\'e, CEA, CNRS, AIM, 91191, Gif-sur-Yvette, France\label{aff83}
\and
Institut de F\'{i}sica d'Altes Energies (IFAE), The Barcelona Institute of Science and Technology, Campus UAB, 08193 Bellaterra (Barcelona), Spain\label{aff84}
\and
DARK, Niels Bohr Institute, University of Copenhagen, Jagtvej 155, 2200 Copenhagen, Denmark\label{aff85}
\and
Waterloo Centre for Astrophysics, University of Waterloo, Waterloo, Ontario N2L 3G1, Canada\label{aff86}
\and
Department of Physics and Astronomy, University of Waterloo, Waterloo, Ontario N2L 3G1, Canada\label{aff87}
\and
Perimeter Institute for Theoretical Physics, Waterloo, Ontario N2L 2Y5, Canada\label{aff88}
\and
Centre National d'Etudes Spatiales -- Centre spatial de Toulouse, 18 avenue Edouard Belin, 31401 Toulouse Cedex 9, France\label{aff89}
\and
Institute of Space Science, Str. Atomistilor, nr. 409 M\u{a}gurele, Ilfov, 077125, Romania\label{aff90}
\and
Consejo Superior de Investigaciones Cientificas, Calle Serrano 117, 28006 Madrid, Spain\label{aff91}
\and
Universidad de La Laguna, Dpto. Astrof\'\i sica, E-38206 La Laguna, Tenerife, Spain\label{aff92}
\and
Dipartimento di Fisica e Astronomia "G. Galilei", Universit\`a di Padova, Via Marzolo 8, 35131 Padova, Italy\label{aff93}
\and
Institut f\"ur Theoretische Physik, University of Heidelberg, Philosophenweg 16, 69120 Heidelberg, Germany\label{aff94}
\and
Universit\'e St Joseph; Faculty of Sciences, Beirut, Lebanon\label{aff95}
\and
Departamento de F\'isica, FCFM, Universidad de Chile, Blanco Encalada 2008, Santiago, Chile\label{aff96}
\and
Universit\"at Innsbruck, Institut f\"ur Astro- und Teilchenphysik, Technikerstr. 25/8, 6020 Innsbruck, Austria\label{aff97}
\and
Satlantis, University Science Park, Sede Bld 48940, Leioa-Bilbao, Spain\label{aff98}
\and
Department of Physics, Royal Holloway, University of London, Surrey TW20 0EX, UK\label{aff99}
\and
Instituto de Astrof\'isica e Ci\^encias do Espa\c{c}o, Faculdade de Ci\^encias, Universidade de Lisboa, Tapada da Ajuda, 1349-018 Lisboa, Portugal\label{aff100}
\and
Cosmic Dawn Center (DAWN)\label{aff101}
\and
Niels Bohr Institute, University of Copenhagen, Jagtvej 128, 2200 Copenhagen, Denmark\label{aff102}
\and
Universidad Polit\'ecnica de Cartagena, Departamento de Electr\'onica y Tecnolog\'ia de Computadoras,  Plaza del Hospital 1, 30202 Cartagena, Spain\label{aff103}
\and
Infrared Processing and Analysis Center, California Institute of Technology, Pasadena, CA 91125, USA\label{aff104}
\and
Dipartimento di Fisica e Scienze della Terra, Universit\`a degli Studi di Ferrara, Via Giuseppe Saragat 1, 44122 Ferrara, Italy\label{aff105}
\and
Istituto Nazionale di Fisica Nucleare, Sezione di Ferrara, Via Giuseppe Saragat 1, 44122 Ferrara, Italy\label{aff106}
\and
INAF, Istituto di Radioastronomia, Via Piero Gobetti 101, 40129 Bologna, Italy\label{aff107}
\and
Astronomical Observatory of the Autonomous Region of the Aosta Valley (OAVdA), Loc. Lignan 39, I-11020, Nus (Aosta Valley), Italy\label{aff108}
\and
Universit\'e C\^{o}te d'Azur, Observatoire de la C\^{o}te d'Azur, CNRS, Laboratoire Lagrange, Bd de l'Observatoire, CS 34229, 06304 Nice cedex 4, France\label{aff109}
\and
Department of Physics, Oxford University, Keble Road, Oxford OX1 3RH, UK\label{aff110}
\and
Aurora Technology for European Space Agency (ESA), Camino bajo del Castillo, s/n, Urbanizacion Villafranca del Castillo, Villanueva de la Ca\~nada, 28692 Madrid, Spain\label{aff111}
\and
Zentrum f\"ur Astronomie, Universit\"at Heidelberg, Philosophenweg 12, 69120 Heidelberg, Germany\label{aff112}
\and
ICL, Junia, Universit\'e Catholique de Lille, LITL, 59000 Lille, France\label{aff113}
\and
ICSC - Centro Nazionale di Ricerca in High Performance Computing, Big Data e Quantum Computing, Via Magnanelli 2, Bologna, Italy\label{aff114}
\and
Instituto de F\'isica Te\'orica UAM-CSIC, Campus de Cantoblanco, 28049 Madrid, Spain\label{aff115}
\and
CERCA/ISO, Department of Physics, Case Western Reserve University, 10900 Euclid Avenue, Cleveland, OH 44106, USA\label{aff116}
\and
Technical University of Munich, TUM School of Natural Sciences, Physics Department, James-Franck-Str.~1, 85748 Garching, Germany\label{aff117}
\and
Max-Planck-Institut f\"ur Astrophysik, Karl-Schwarzschild-Str.~1, 85748 Garching, Germany\label{aff118}
\and
Laboratoire Univers et Th\'eorie, Observatoire de Paris, Universit\'e PSL, Universit\'e Paris Cit\'e, CNRS, 92190 Meudon, France\label{aff119}
\and
Departamento de F{\'\i}sica Fundamental. Universidad de Salamanca. Plaza de la Merced s/n. 37008 Salamanca, Spain\label{aff120}
\and
Universit\'e de Strasbourg, CNRS, Observatoire astronomique de Strasbourg, UMR 7550, 67000 Strasbourg, France\label{aff121}
\and
Center for Data-Driven Discovery, Kavli IPMU (WPI), UTIAS, The University of Tokyo, Kashiwa, Chiba 277-8583, Japan\label{aff122}
\and
Dipartimento di Fisica - Sezione di Astronomia, Universit\`a di Trieste, Via Tiepolo 11, 34131 Trieste, Italy\label{aff123}
\and
California Institute of Technology, 1200 E California Blvd, Pasadena, CA 91125, USA\label{aff124}
\and
Department of Physics \& Astronomy, University of California Irvine, Irvine CA 92697, USA\label{aff125}
\and
Kapteyn Astronomical Institute, University of Groningen, PO Box 800, 9700 AV Groningen, The Netherlands\label{aff126}
\and
Departamento F\'isica Aplicada, Universidad Polit\'ecnica de Cartagena, Campus Muralla del Mar, 30202 Cartagena, Murcia, Spain\label{aff127}
\and
Instituto de F\'isica de Cantabria, Edificio Juan Jord\'a, Avenida de los Castros, 39005 Santander, Spain\label{aff128}
\and
INFN, Sezione di Lecce, Via per Arnesano, CP-193, 73100, Lecce, Italy\label{aff129}
\and
Department of Mathematics and Physics E. De Giorgi, University of Salento, Via per Arnesano, CP-I93, 73100, Lecce, Italy\label{aff130}
\and
INAF-Sezione di Lecce, c/o Dipartimento Matematica e Fisica, Via per Arnesano, 73100, Lecce, Italy\label{aff131}
\and
CEA Saclay, DFR/IRFU, Service d'Astrophysique, Bat. 709, 91191 Gif-sur-Yvette, France\label{aff132}
\and
Institute of Cosmology and Gravitation, University of Portsmouth, Portsmouth PO1 3FX, UK\label{aff133}
\and
Department of Computer Science, Aalto University, PO Box 15400, Espoo, FI-00 076, Finland\label{aff134}
\and
Instituto de Astrof\'{\i}sica de Canarias, E-38205 La Laguna; Universidad de La Laguna, Dpto. Astrof\'\i sica, E-38206 La Laguna, Tenerife, Spain\label{aff135}
\and
Caltech/IPAC, 1200 E. California Blvd., Pasadena, CA 91125, USA\label{aff136}
\and
Ruhr University Bochum, Faculty of Physics and Astronomy, Astronomical Institute (AIRUB), German Centre for Cosmological Lensing (GCCL), 44780 Bochum, Germany\label{aff137}
\and
Department of Physics and Astronomy, Vesilinnantie 5, University of Turku, 20014 Turku, Finland\label{aff138}
\and
Serco for European Space Agency (ESA), Camino bajo del Castillo, s/n, Urbanizacion Villafranca del Castillo, Villanueva de la Ca\~nada, 28692 Madrid, Spain\label{aff139}
\and
ARC Centre of Excellence for Dark Matter Particle Physics, Melbourne, Australia\label{aff140}
\and
Centre for Astrophysics \& Supercomputing, Swinburne University of Technology,  Hawthorn, Victoria 3122, Australia\label{aff141}
\and
Department of Physics and Astronomy, University of the Western Cape, Bellville, Cape Town, 7535, South Africa\label{aff142}
\and
Universit\'e Libre de Bruxelles (ULB), Service de Physique Th\'eorique CP225, Boulevard du Triophe, 1050 Bruxelles, Belgium\label{aff143}
\and
DAMTP, Centre for Mathematical Sciences, Wilberforce Road, Cambridge CB3 0WA, UK\label{aff144}
\and
Kavli Institute for Cosmology Cambridge, Madingley Road, Cambridge, CB3 0HA, UK\label{aff145}
\and
Department of Astrophysics, University of Zurich, Winterthurerstrasse 190, 8057 Zurich, Switzerland\label{aff146}
\and
Department of Physics, Centre for Extragalactic Astronomy, Durham University, South Road, Durham, DH1 3LE, UK\label{aff147}
\and
IRFU, CEA, Universit\'e Paris-Saclay 91191 Gif-sur-Yvette Cedex, France\label{aff148}
\and
Univ. Grenoble Alpes, CNRS, Grenoble INP, LPSC-IN2P3, 53, Avenue des Martyrs, 38000, Grenoble, France\label{aff149}
\and
Dipartimento di Fisica, Sapienza Universit\`a di Roma, Piazzale Aldo Moro 2, 00185 Roma, Italy\label{aff150}
\and
Centro de Astrof\'{\i}sica da Universidade do Porto, Rua das Estrelas, 4150-762 Porto, Portugal\label{aff151}
\and
Instituto de Astrof\'isica e Ci\^encias do Espa\c{c}o, Universidade do Porto, CAUP, Rua das Estrelas, PT4150-762 Porto, Portugal\label{aff152}
\and
HE Space for European Space Agency (ESA), Camino bajo del Castillo, s/n, Urbanizacion Villafranca del Castillo, Villanueva de la Ca\~nada, 28692 Madrid, Spain\label{aff153}
\and
INAF - Osservatorio Astronomico d'Abruzzo, Via Maggini, 64100, Teramo, Italy\label{aff154}
\and
Theoretical astrophysics, Department of Physics and Astronomy, Uppsala University, Box 516, 751 37 Uppsala, Sweden\label{aff155}
\and
Mathematical Institute, University of Leiden, Einsteinweg 55, 2333 CA Leiden, The Netherlands\label{aff156}
\and
Institute of Astronomy, University of Cambridge, Madingley Road, Cambridge CB3 0HA, UK\label{aff157}
\and
Department of Astrophysical Sciences, Peyton Hall, Princeton University, Princeton, NJ 08544, USA\label{aff158}
\and
Space physics and astronomy research unit, University of Oulu, Pentti Kaiteran katu 1, FI-90014 Oulu, Finland\label{aff159}
\and
Institut de Physique Th\'eorique, CEA, CNRS, Universit\'e Paris-Saclay 91191 Gif-sur-Yvette Cedex, France\label{aff160}
\and
Center for Computational Astrophysics, Flatiron Institute, 162 5th Avenue, 10010, New York, NY, USA\label{aff161}}

\abstract{We compare the performance of the flat-sky approximation and Limber approximation for the clustering analysis of the photometric galaxy catalogue of \Euclid. We study a 6 bin configuration representing the first data release (DR1) and a 13 bin configuration representative of the third and final data release
(DR3). We find that the Limber approximation is sufficiently accurate for the analysis of the wide bins of  DR1. Contrarily, the 13 bins of DR3 cannot be modelled accurately with the Limber approximation.  Instead, the flat-sky approximation is accurate to below $5\%$ in recovering the angular power spectra of galaxy number counts in both cases and can be used to simplify the computation of the full power spectrum in harmonic space for the data analysis of DR3.}


   \titlerunning{Flat-sky approximation for \Euclid}
   \authorrunning{Euclid Collaboration: W.L.\ Matthewson et al. }
   
   \maketitle
   
\section{\label{sc:Intro}Introduction}
The \Euclid satellite will generate two main surveys. A survey of about $10^7$ galaxies with very precise spectroscopic redshifts and a second survey of about $1.5\times10^9$  galaxies with less precise photometric redshifts, but which contains also galaxy shapes for a weak lensing shear analysis \citep{EuclidSkyOverview}. One of the main summary statistics of this second survey will be the 3\texttimes2 angular power spectra of the lensing shear, the galaxy number counts, and their cross correlation \citep{Blanchard-EP7,Tutusaus20}.

In order to be able to exploit the data optimally, the calculation of the theoretical power spectra to be compared with the data must be both fast and accurate. For example, when performing an MCMC analysis for cosmological parameter inference, many theoretical spectra will need to be calculated, as quickly as possible, without biasing results, something which is currently not feasible to do by using the full calculation. So far, mainly the Limber approximation (which can provide a speed up on the order of $100$ times with respect to the full calculation in \texttt{CLASS}, for example) is used to determine power spectra from photometric surveys \citep[for a recent topical paper, see][]{2023OJAp....6E...8L}. While the Limber approximation is excellent when determining the power spectrum of the lensing potential via shear measurements~\citep[][]{Kitching:2016zkn,Kilbinger:2017lvu}, it is well-known to perform poorly for galaxies distributed narrowly in the radial direction \citep[e.g.][]{2007A&A...473..711S}.  In this paper we aim to assess the particular impact of this failure on the results set to come from \Euclid, both in the first and final data releases. This is done by determining the errors incurred, relative to the full calculation, by the Limber approximation and another approximation, the `flat-sky approximation', which improves accuracy at large scales, while also speeding up the computation with respect to the full calculation, if using an optimised code. While it does not use the flat sky approximation, \texttt{BLAST} (Beyond Limber Angular power Spectra Toolkit) is one such example of a highly optimised code for calculating the power spectrum \citep{Chiarenza:2024rgk}.

The paper is structured as follows. In the next section we present a brief introduction to the flat-sky approximation and the Limber approximation and make contact with the relevant literature.
In Sect.~\ref{sc:num} we discuss the numerical implementation of the approximations used in this work, and in Sect.\ \ref{sc:res} we present results for the photometric survey of \Euclid. In Sect.\ \ref{sc:con} we discuss our results.
\vspace{0.2cm}

\paragraph{Notation.}
We shall calculate angular power spectra of different variables in different redshift bins. Power spectra of variables $A$ and $B$ at redshifts $z$ and $z'$ calculated with the full expression (by using numerical codes like \texttt{CAMB}, \citealt{Challinor:2011bk}, or \texttt{CLASS}, \citealt{DiDio:2013bqa}) will be called $C^{AB}_\ell(z,z')$ or, when integrated over the redshift bins $i$ and $j$, $C^{AB}_\ell(i,j)$. When neither $A$ nor $B$ are present in the superscript, $C_\ell(i,j)$ refers to the total observed power spectrum, including correlations of all necessary variables. The corresponding notations for the flat-sky approximation and Limber approximation will be $^{\rm F}C^{AB}_\ell(z,z')$ and $^{\rm L}C^{AB}_\ell(z,z')$ or, when integrated over redshift bins $i$ and $j$, we denote them $^{\rm F}C^{AB}_\ell(i,j)$ and $^{\rm L}C^{AB}_\ell(i,j)$, respectively.

We work in a spatially flat background Friedmann-Lema\^\i tre universe with conformal time denoted $\eta$. As spatial curvature is certainly very small \citep[see][]{Aghanim:2018eyx} this is sufficient for our purpose. While it is fairly straightforward to extend the expressions for the angular power spectrum of galaxy number counts and its approximations to curved cosmologies, the expressions are more complicated and beyond the scope of our study. The background metric is then given by
\be
\ud s^2= a^2(\eta)\,\left[-c^2\ud\eta^2+\ud\vec x\cdot\ud\vec x\right] \;,
\ee
with $c$, the vacuum speed of light. The scale factor is normalized to one today, $a(\eta_0)\equiv a_0=1$. The Hubble factor is defined as $H = \frac{1}{a}\frac{{\rm d}a}{{\rm d}t}$, where $t$ (${\rm d}t = a\,{\rm d}\eta$) is the cosmic time.

\section{\label{sc:flat} The flat-sky and Limber approximations for the angular power spectrum of galaxy number counts}

The flat-sky approximation has been investigated in the past \citep[see][]{Datta:2006vh,White:2017ufc,Castorina:2018nlb,Jalilvand:2019brk,Matthewson:2020rdt,Gao:2023tcd}. While it has been found to be in excellent agreement with the full calculation of angular power spectra for very slim redshift bins \citep[see][]{Matthewson:2020rdt}, an accurate extension to photometric bins has been developed in \citet{Gao:2023tcd}. Here we present this extension (the flat-sky approximation) along with a second approach (the Limber approximation) as means to approximate the angular power spectrum for an arbitrary observable in the sky \citep[see also][]{EP-Tessore}. We then apply {both these approximations} to the angular power spectrum of galaxy number counts.

Let us first consider an arbitrary cosmological observable $A(\bx,\eta)$ at redshift $z$ which we observe in the sky in direction $\bn$. If $r(z)$ is the comoving distance out to redshift $z$, in terms of the variables $\bn$ and $z$, we have
\be\label{e:Asimple}
A(\bn,z) = A\big(r(z)\,\bn,\eta_0-r(z)/c\big)\;,
\ee
where $\eta_0$ is the present time. We typically expand this function on the sphere using spherical harmonics (if $A$ is a complex spin-$s$ field, like the shear, we use spin-weighted spherical harmonics),
\be
A(\bn,z) = \sum_{\ell,m}a_{\ell m}(z)\,Y_{\ell m}(\bn)\;.
\ee
Assuming statistical homogeneity and isotropy, only expectation values of equal $\ell$ and $m$ do not vanish and their value depends only on $\ell$. These define the angular power spectrum of two variables $A(\bn,z)$ and $B(\bn',z')$, viz.\
\be
\langle a^A_{\ell m}(z)\,a^{B^*}_{\ell' m'}(z')\rangle =\de^{\rm K}_{\ell\ell'}\,\de^{\rm K}_{mm'}\,C^{AB}_\ell(z,z')\;.
\label{e:statis}
\ee
For $A=B$, we talk of the auto-correlation power spectrum, which is always positive definite, while for $A\neq B$ we have a cross-correlation spectrum. If $A(\bn,z)$ is of the simple form of \cref{e:Asimple}, we can express its angular power spectrum in terms of the 3D power spectrum in Fourier space as \citep[see, e.g.][]{Durrer:2020fza}
\be\label{e:CAlzz'}
C^{AB}_\ell(z,z') = \frac2\pi\,\int_0^\infty\udln k\,k^3\,P_{AB}(k,z,z')\,j_\ell(k\,r)\,j_\ell(k\,r') \;.
\ee
Here, $r$ and $r'$ are the comoving distances out to redshifts $z$ and $z'$, respectively, and $j_\ell$ is the spherical Bessel function of order $\ell$, whilst $P_{AB}(k,z,z')$ is the 3D unequal-time power spectrum of $A$ and $B$, namely
\be
\langle A(\bk,z)\,B^*(\bk',z')\rangle = (2\,\pi)^3\,\de^{\rm D}(\bk-\bk')\,P_{AB}(k,z,z') \;.
\ee

If the variables $A$ or $B$ also have an intrinsic dependence on the direction $\bn$, the spherical Bessel functions in \cref{e:CAlzz'} may be replaced by more complicated expressions. For example, for redshift space distortions we find ${{\rm d}^2j_\ell}/{{\rm d}x^2}$ instead of $j_\ell(x)$ \citep[see][for details on all terms appearing in the number counts]{Bonvin:2011bg}. Henceforth, we shall refer to various terms with $A,B$ being replaced with the shorthand: D for density, RSD for redshift space distortions, and L for lensing.

More realistically, especially for photometric surveys, we do not measure the $C_\ell$ at a precise redshift but integrated over redshift windows, $w_i(z)$.
\Cref{e:CAlzz'} then becomes 
\be
C^{AB}_\ell(i,j) = \int_0^\infty\ud z\,\ud z'\,w_i(z) \,w_j(z')\,C^{AB}_\ell(z,z')\;,
\label{e:CAlijfull}
\ee
where, depending on the observable, $j_\ell$ in $C^{AB}_\ell(z,z')$ may have to be replaced by its first or second derivative, as noted before in the case of redshift space distortions.

The Limber approximation now consists of replacing the integral over $k$ in \cref{e:CAlzz'} by a Dirac-delta function such that \citep{Limber:1954zz,1992ApJ...388..272K}
\begin{align}
C^{AB}_\ell(z,z') &\simeq \frac{\de^{\rm D}(r-r')}{r^2}\,P_{AB}\left(\frac{\ell+1/2}{\ r},z,z\right) \nonumber \\
&=\de^{\rm D}(z-z')\,\frac{H(z)}{cr^2}\,
P_{AB}\left(\frac{\ell+1/2}{r},z,z\right)\;,
\end{align}
hence
\be\label{e:CAlijLimber}
 ^{\rm L}C^{AB}_\ell(i,j) \simeq \int_0^\infty
\ud z\,w_i(z)\,w_j(z)\,\frac{H(z)}{cr^2(z)}\,P_{AB}\left(\frac{\ell+1/2}r,z,z\right)  \;.
\ee
In this way, the heavily oscillating integral over $k$ and the double integral over $z$ and $z'$ can be replaced by a single integral  over $z$. If $A=B$, the integrand is even a manifestly non-negative function. This provides an enormous speed-up leading to excellent results for weak lensing which has a very broad window function at multipoles $\ell\gtrsim 20$ \citep[]{Kilbinger:2017lvu}. However, it is well known that for galaxy number counts, especially in slim redshift bins, the Limber approximation differs by a factor 2 and more (depending on the bin width) from the true result for $\ell\lesssim 100$ \citep[][]{DiDio:2014lka,DiDio:2018unb,Fang:2019xat,2022MNRAS.510.1964M}. Here we compare it with the flat-sky approximation for the photometric survey of \Euclid.

{Let us now introduce the flat-sky approximation: it} approximates the sky by a plane and the pair $(\ell,m)$ is replaced by a 2D vector $\bel$. We assume that $\bn$ is close to some reference direction $\bde$, and we write $\bn=\bde+\bal$. The amplitudes $a_{\ell m}$ are then replaced by 2D Fourier transforms,
\begin{align}\label{e:aflat}
a^A(\bel,z) &= \frac{1}{2\,\pi}\,\int_{\real^2}
\ud^2\al\, A(\bal,z)\,\er^{-\ii\,\bal\cd\bel}\;, \\
A(\bal,z) &= \frac{1}{2\,\pi}\,
\int_{\real^2}\ud^2\ell\, a^A(\bel,z)\,\er^{\ii\,\bal\cd\bel}\;.
\end{align}
Inserting the 3D Fourier representation for $A$, we find
\be\label{e:A3d}
A(\bal,z) =\frac 1{(2\,\pi)^3}\,\int_{\real^3}\ud^3k\,A(\bk,z)\,\er^{\ii\,\bk\cd(\bde+\bal)\,r(z)}\;.
\ee
Setting
\begin{align}
\bk &= k_\|\,\bde + \bk_\perp
\label{eq:k_of_l}
\end{align}
and comparing \cref{e:A3d,e:aflat}, we obtain
\begin{align}
a^A(\bel,z) &=\frac 1{(2\,\pi)^2}\,\int_{\real^3}\ud^3k\,A(\bk,z)\,\er^{\ii\,\bk\cd(\bde+\bal)\,r(z)}\nonumber
\de^{\rm D}[\bel-r(z)\,\bk_\perp] \\
&= \frac 1{(2\,\pi)^2\,r^2(z)}
\int_{-\infty}^{+\infty}\ud k_{\|}\,A(\bk(k_\|,\ell,z),z)\,\er^{\ii\,k_{\|}\,r(z)}\;,
\label{e:alflat}
\end{align}
where $\bk(k_\|,\ell,z)$ is given by \cref{eq:k_of_l} with $\bk_\perp = \bel/r(z)$. 
This expression for $\bk_\perp$ is reminiscent of the Limber approximation, but overall the flat-sky approximation is distinct, since we retain the parallel component of $\bk$.
If the redshifts are similar, we may replace $z$ and $z'$ by $\bar z$, defined by $r(\bar z) = \sqrt{r(z)r(z')}$, in $\bel/r(z)$, such that
\begin{multline}
 \langle a^A(\bel,z)\,a^{B*}(\bel',z')\rangle \simeq\frac1{2\,\pi} \\
 \times\int_{\real^3}{\ud^3k}\,P_{AB}(k,\bar z)\,\er^{\ii\,k_\|\,(r -r')}\,\de^{\rm D}[\bel-r(\bar z)\,\bk_\perp]\,\de^{\rm D}[\bel'-r(\bar z)\,\bk_\perp] \\
= \de^{\rm D}(\bel-\bel')\,\frac 1{2\,\pi\,r^2(\bar z)}\,\int_{-\infty}^\infty\ud k_\|\,P_{AB}(k,\bar z,\bar z)\,\er^{\ii\,k_\|\,(r -r')} \;,
\end{multline}
in which, {the imaginary part vanishes and  the real part, gives}
\be \label{e:CAlflat}
^{\rm F}C^{AB}_\ell(z,z') = \frac 1{\pi\,r^2(\bar z)}\,\int_{0}^\infty\ud k_\|\,P_{AB}(k,\bar z,\bar z)\,\cos[k_\|\,(r -r')] \;.
\ee

Note that we would not recover
the physically relevant factor $\de^{\rm D}(\bel-\bel')$, were it not for the fact that we had set $r=r(\bar z)=r'$ in the expression for $\bk_\perp$. This is due to the fact that $A(\bal,z)$ and 
$A(\bal,z')$ live on spheres with different radii, which modifies the relation between $\bk_\perp$ and $\bel$. On the other hand, in the full calculation, see \cref{e:statis,e:CAlijfull}, statistical isotropy ensures $\de^{\rm K}_{\ell\,\ell'}$ and $\de^{\rm K}_{m\,m'}$. Fortunately, the oscillations of the cosine in \cref{e:CAlflat} significantly suppress the signal for different redshifts so that this inaccuracy is irrelevant because of the shape of the power spectrum, which decreases in amplitude at small $k$, see \cite{Gao:2023tcd, Gao_2023}.

The power spectra from redshift bins with window functions $w_i$ and $w_j$ are now given by
\begin{align}
^{\rm F}C^{AB}_\ell(i,j) &\simeq\int_0^\infty\frac{\ud z\,\ud z'}{\pi\,r^2(\bar z)}\,w_i(z)\,w_j(z')\nonumber\\
&\quad\times\int_0^\infty \ud k_\|\,P_{AB}\left(k,\bar z,\bar z\right)\,\cos[k_\|\,(r -r')]  \;, \label{e:CAlijflat2}\\
&= \int_0^\infty \frac{\ud z\,\ud z'}{\pi\,r^2(\bar z)}\,w_i(z)\,w_j(z')\nonumber\\
&\quad\times\int_{\ell/r(\bar z)}^\infty \ud k\,P_{AB}\left(k,\bar z,\bar z\right)\,\cos[k_\|\,(r -r')]  \;,
\label{e:flatskyfin}
\end{align}
where the change in integral bounds follow from the change of variable, given the relation between $k$ and $k_\|$, 
\be\label{e:kofl}
k =\sqrt{k^2_\| +\frac{\ell^2}{r^2(\bar z)}} \;.
\ee
In the literature, including in \citet{Gao:2023tcd}, the significantly better `recalibrated' flat-sky
approximation is generally employed,
where $\ell$ is replaced by $\ell+1/2$ in the above expression for $k$.

Even though the number of integrals over the power spectrum $P_{AB}\left(k,z,z'\right)$, for the flat-sky expression in \cref{e:flatskyfin}, is the same as in \cref{e:CAlijfull}, for $r=r'$ (the value with the dominant contribution to $C^{AB}_{ij}$) the $\ell$-dependence has simply been absorbed into the boundary of the integral, reducing the complexity of the integration. In addition, the integral no longer contains the heavily oscillating product of spherical Bessel functions. More importantly, as described in \citet{Gao:2023tcd}, this form of the integral  allows the $k$-integral to be precomputed and stored, which significantly saves computation time in applications where the angular power spectra from many combinations of different cosmological parameter values are required, such as parameter estimation and MCMC analysis. From the flat-sky approximation we may further obtain the Limber expression simply by setting $k_\|=0$ and replacing the $k$-integral by $\de^{\rm D}(r-r')=\de^{\rm D}(z-z')H(z)/c$.

We shall use this flat-sky expression to calculate the $C_\ell$ for galaxy number counts. In addition to density fluctuations, we also include redshift space distortions and relativistic effects, for example, the lensing magnification, which arise in observations made on the past light cone -- see \citet{2009PhRvD..80h3514Y}, \citet{Bonvin:2011bg}, and \citet{Challinor:2011bk} for a full derivation of all terms in the exact solution. We refer to these as the contributions from observational effects.
For the density we have
\be\label{e:dens}
A^{\rm D}(\bk,z) = b(z)\,D(\bk,z) \;,
\ee
where $b(z)$ is the linear galaxy bias, that is determined via simulations (see Sect.~\ref{s:flag}), and $D(\bk,z)$ is the matter density contrast in Fourier space. The redshift space distortions are given by
\be\label{e:reds}
A^{\rm RSD}(\bk,z)= 
\frac{k^2_\|}{k\,\HH(z)}\,V(\bk,z) \;,
\ee
where $k^{-1}\,V(\bk,z)$ is the Fourier transform of the velocity potential, and $\HH = \frac{1}{a}\frac{{\rm d}a}{{\rm d}\eta} = H\,a$ is the conformal Hubble factor. The next important contribution to the number counts comes from the lensing magnification and is given by
\be
A^{\rm L}(\bk,z,L) = \Big( 2-5\,s(z,L)\Big)\,\kappa \,,
\ee
where $\ka$ is the lensing convergence and contains an integral over redshift, which is well approximated by the Limber approximation.
The function $s(z,L)$ is the logarithmic slope of the galaxy number density, $n(z,L)$, as a function of luminosity, $L$,
\be\label{e:magbias}
s(z,L) = \frac{2}{5}\frac{\dd \ln n(z,L)}{\dd \ln L}\,.
\ee
We shall use the result obtained by the Flagship simulation (see Sect.~\ref{s:flag}).
 For a more detailed representation of all Limber and flat-sky expressions and their cross correlations, see \citet{Matthewson:2020rdt}. In this analysis we include the appropriate terms for these three contributions in each of the approximations of the full angular power spectra. The remaining relativistic contributions in \citet[]{Bonvin:2011bg} are negligible and, therefore, set to zero in both approximations.

\section{Implementation, performance, and simulations \label{sc:num}}
In this section we first describe the implementation of our power spectra computations under different approximations. We then clarify the methodology that has been used to compare them, and finally present the different \Euclid scenarios that were studied in this work.

\subsection{Computation of the power spectra}
As laid out in the previous section, we consider three alternative methods to compute the power spectra of galaxy number counts; that is, the full computation~(Eq.\,\ref{e:CAlijfull}), the flat-sky approximation~(Eq.\,\ref{e:flatskyfin}), and the Limber approximation~(Eq.\,\ref{e:CAlijLimber}). Starting with the full computation case, we 
calculate the galaxy number count angular power spectra with the \texttt{CLASS} Boltzmann solver \citep{2011JCAP...07..034B} to linear order, including all contributions and relativistic terms. For the flat-sky approximation, we make use of the \texttt{GZCphysics} code described in \citet{Gao:2023tcd}, further modified by us to use \texttt{CLASS} instead of the Boltzmann solver \texttt{CAMB}. We  checked that the flat-sky results from the above paper are in agreement with our analogous calculations.
We then extend the code to include lensing magnification, and to accept arbitrary (not just Gaussian) windows in redshift, needed for the application to \Euclid. Following this, we implement the \Euclid photometric survey simulated tomographic bins. Finally, we compute the power spectra with the Limber approximation using \texttt{CAMB} via the \texttt{CosmoSIS} framework \citep{Zuntz:2014csq}. This choice is made to {facilitate the control of the scale at which the Limber approximation is used in the computation which is not straightforward when} simply using the Limber implementation available in \texttt{CLASS} or \texttt{CAMB}.

Our main goal in this analysis is to compare the performance of the flat-sky approximation and Limber approximation with respect to the full computation. Given that the differences between the approximations and the full result are only relevant at low multipoles, we restrict our analysis to the largest scales ($\ell\leq300$) and consider a linear matter power spectrum. We note that even if there is some leakage from higher $k$ modes due to integrated terms like lensing magnification, when applying a multipole cut, there is no need to consider accurate recipes to account for the non-linearities in the matter power spectrum at the scales relevant for this analysis.

\subsection{\Euclid specifications}
In this work we extract the specifications that correspond to the \Euclid photometric survey from the Flagship simulation for two different scenarios: the first data release (DR1) and the third and final data release (DR3). {Below, both the simulation and the data releases are briefly described.}

\subsubsection{The Flagship simulation}\label{s:flag}
The Flagship galaxy mock \citep{EuclidSkyFlagship} is a simulated catalogue of galaxies built from one of the largest $N$-body dark matter simulations ever run, with a box size of $3600\,h^{-1}\,\mathrm{Mpc}$ and a particle mass resolution of $10^9\,h^{-1}\,M_{\odot}$. The dark matter simulation was run with \texttt{PKDGRAV3} \citep{Potter:2016ttn}, and haloes were identified with the \texttt{rockstar} algorithm \citep{2013ApJ...762..109B}. Haloes were then populated with galaxies using a combination of the halo occupation distribution and sub-halo abundance matching techniques \citep{2015MNRAS.447..646C}. The galaxy mock was calibrated to reproduce several observations, including the luminosity function\,\citep{Blanton2003b,Blanton2005b}, the clustering of galaxies\,\citep{Zehavi2011}, and the colour-distribution as a
function of the absolute magnitude in the $r$ band\,\citep{Blanton2005c} of Sloan Digital Sky Survey data.

The cosmological parameters for this simulation are given in \cref{table:cosmo} along with the ones used in this paper, which are quite similar, but not exactly the same. Since the only simulation output used here is the galaxy distribution as a function of redshift in each tomographic bin,
 these small differences in cosmological parameters are not relevant. What is relevant is that we use the same galaxy redshift distribution, linear galaxy bias, and magnification bias for all three calculations. The catalogue can be accessed via the CosmoHub platform\footnote{\url{https://cosmohub.pic.es/home}} \citep{cosmohub_ref1,cosmohub_ref2}.

\subsubsection{DR1 settings}
In order to study the performance of the flat-sky approximation and Limber approximation for \Euclid, we first consider the initial data release. We follow the approach presented in \citet{EuclidSkyOverview} and select a galaxy sample from the Flagship galaxy mock, selecting a conservative magnitude cut at $\IE<23.5$. This choice of a fairly bright sample decreases the presence of systematic effects that introduce spurious clustering signals. Once the sample has been defined, we classify galaxies into six equi-populated bins, depending on the photometric redshift assigned to each galaxy. The galaxy distributions are then built by making a histogram using the bins in observed redshift associated to each object. We note that these are derived assuming Legacy Survey of Space and Time (LSST)-like photometry, which will not be available for DR1 and, as such, these DR1 settings correspond to an optimistic scenario. We summarise the main specifications in \cref{table:SB}. In this case, we measured the corresponding galaxy bias {defined in \cref{e:dens}, and the magnification bias defined in \cref{e:magbias}} for this sample and provide their values in \cref{table:6bin}. We also show the galaxy distributions in \cref{f:EuclidBins}.

\subsubsection{DR3 settings}
\begin{figure}[ht]
\centering
\includegraphics[width=0.95\columnwidth]{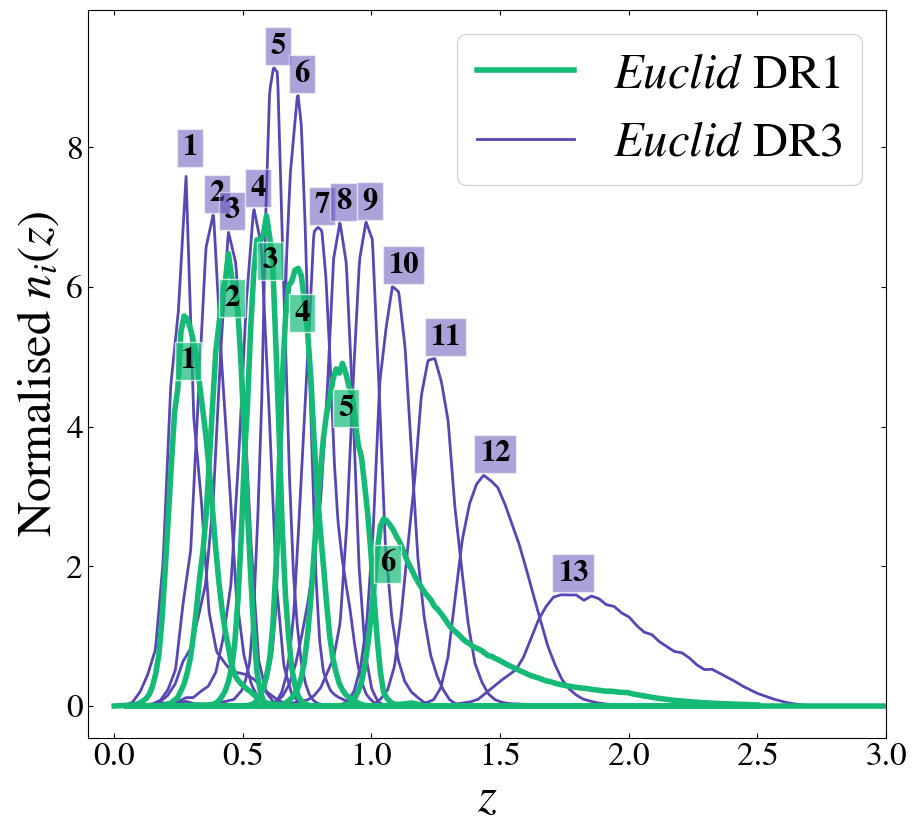}
\caption{The number density bins simulated for the 
photometric surveys of \Euclid, in six bins and 13 bins equi-populated configurations. The numbering convention we use for the $i$th bin is shown in the coloured blocks above the maxima of each bin.} 
\label{f:EuclidBins}
\end{figure}

In addition to the initial \Euclid data release, we also consider the performance of the flat-sky approximation and Limber approximation for the full data sample of \Euclid at the end of the mission. We consider 13 equi-populated bins with a higher (dimmer) magnitude cut at $\IE<24.5$. The final sample contains a total number density of 24.3 galaxies per arcmin$^2$, and we consider the galaxy and magnification biases measured for this sample in \citet{EuclidSkyOverview}. We summarise the main specifications in \cref{table:SB} and show the galaxy distributions in \cref{f:EuclidBins}.

\begin{table}[ht]
\small
\centering
\caption{Cosmological parameters for the Flagship simulation (taken from the \Euclid reference cosmology), and the fiducial cosmology used in the current work, taken from \Planck 2018 (`TT, TE, EE+lowE+lensing+BAO' results, \citealt{Aghanim:2018eyx}).}
\begin{tabular}{lllllll}
&$h$& $\Omega_{\rm c}\,h^2$ & $\Omega_{\rm b}\,h^2$ & $n_{\rm s}$ & $10^9\,A_{\rm s}$ \\
\hline 
Flagship sim.&$0.6700$ &$0.12120$ &$0.02200$ &$0.9600$ &$2.100$ \\
This work&$0.6766$ &$0.11933$ &$0.02242$ &$0.9665$ &$2.105$ 
\end{tabular}
\label{table:cosmo}
\tablefoot{Here the Hubble parameter is $H_0 =h\,100$\,km\,s$^{-1}$\,Mpc$^{-1}$, $\Omega_{\rm c}$ is the dark matter density parameter, $\Omega_{\rm b}$ is the baryon density parameter, $n_{\rm s}$ is the scalar spectral index, and $A_{\rm s}$ is the amplitude of the scalar perturbation spectrum at the pivot scale.}
\end{table}

\begin{table}[ht]
\small
\centering
\caption{Summary of survey specifications used in \texttt{SpaceBorne} to calculate the covariance of each binning configuration \citep{EuclidSkyOverview}. }
\begin{tabular}{lll}
Specification& DR1& DR3\\
\hline
\# galaxies [$\mathrm{arcmin}^{-2}]$& $10.13$ & $24.3$\\
Intrinsic ellipticity dispersion &  \multirow{ 2}{*}{0.26} & \multirow{ 2}{*}{0.26}\\
(per component)& &\\
Survey area [$\deg^2$]& 2500 &  14\,000\\
\end{tabular}
\label{table:SB}
\end{table}

\begin{table}[ht]
\small
\centering
\caption{Linear galaxy bias factor, $b(z)$,  and magnification bias, $s(z,L)$, for the six bin configuration of \Euclid.}
\begin{tabular}{lllllll}
Bin& 1& 2& 3& 4& 5& 6\\
\hline
Bias factor&1.14& 1.20& 1.32& 1.42& 1.52& 1.94\\
Mag.\ bias&0.206& 0.273&0.261& 0.322& 0.471& 0.789
\end{tabular}
\label{table:6bin}
\end{table}

\subsection{Assessment of performance}
\label{sec:assess}

For both survey configurations considered, we wish to assess the accuracy of each of the approximations in recovering the full calculation result, taking into account the uncertainty on the measurement of the power spectra -- which, crucially, is largest at small multipoles, and depends on the survey configuration under consideration.

To compare the accuracy of each approximation in the context of real measurements, we simulate noisy power spectra, based on the full \texttt{CLASS} spectra and the theoretical covariance matrix for the planned tomographic bins. In this way, we obtain a set of simulated mock spectra that encompasses the expected spread of measurements, subject to the relevant observational uncertainties (see \cref{table:SB,table:6bin}). Specifically, we take the values of the full spectrum for some subset of multipoles $\ell$ as the mean, around which we draw $1000$ random noise realisations which are normally distributed with a Gaussian covariance (i.e., we neglect non-Gaussian contributions from the connected four-point function). We use the \texttt{SpaceBorne} code\footnote{\url{https://github.com/davidesciotti/Spaceborne_covg/tree/main}} \citep{EP-Sciotti} to estimate an analytic Gaussian covariance for the different spectra considered computed using the full spectrum calculation from \texttt{CLASS}.
This Gaussian covariance takes the form 
\begin{align}
{\rm Cov}\left[C_\ell^{GG}(i,j)C_{\ell'}^{G'G'}(k,l)\right] = \frac{1}{2\ell(\ell+1)f_{\rm sky}\Delta\ell}\delta^K_{\ell\ell'}\hspace{45pt}&\\\nonumber
\hspace{20pt}\times\left[\left(C_{\ell}^{GG'}(i,k)+N_{\ell}^{GG'}(i,k)\right)\left(C_{\ell'}^{GG'}(j,l)+N_{\ell'}^{GG'}(j,l)\right)\right.
\hspace{10pt}&\\\nonumber
\hspace{20pt}+\left.\left(C_{\ell}^{GG'}(i,l)+N_{\ell}^{GG'}(i,l)\right)\left(C_{\ell'}^{GG'}(j,k)+N_{\ell'}^{GG'}(j,k)\right)\right]\,,&
\label{e:cov}
\end{align}
see equation (138) of \citet{Euclid:2019clj}, where $f_{\rm sky}$ is the fraction of the sky observed, $\Delta\ell$ is the width of the multipole bins, $\delta^K_{\ell\ell'}$ is the Kronecker delta, $G$ indicates the full observed galaxy clustering power spectrum, and indices $i,j,k,l$ run over the tomographic bins. The noise term is given by the shot noise
\be
N_\ell^{GG}(i,j) = \frac{1}{\bar n_i}\delta^K_{ij}\,,
\ee
with $\bar n_i$ the galaxy surface density per bin in units of inverse steradians.

By comparing to each other the $\chi^2$ distributions, formed from the residuals of each method's result with respect to every one of the spectra in the set of mock observations, we can then obtain an estimate of how far each of the calculations might effectively lie from an observed spectrum, and the frequency with which this occurs. 

The $\chi^2$ for a particular approximation of the spectra ${}^XC_{\ell}$, $X\in\{{\rm L},{\rm F}\}$, with respect to one of the noisy realisations, $\hat{C}_{\ell}$, is computed as
\be
\chi^2_X = \sum_{n=1}^{12} \left(\vec r_n^X\right)^{\sf T}\,(\tens\Sigma^{-1})_n\,\vec r_n^X\;,\label{eq:X2}
\ee
where the sum is over values corresponding to 12 bins distributed evenly in the logarithmic space of $\ell$, between $\ell_{\min}=2$ and $\ell_{\max}=300$.
In \cref{eq:X2}, $\vec r_n^X$ is the vector of the residuals,
\be
\vec r_n^X \coloneqq{\rm vech}_{ij}\left[ \hat{C}_n(i,j)-{}^XC_n(i,j)\right]\;,
\ee
with ${\rm vech}_{ij}$ the half-vectorisation operator\footnote{The vectorisation of a matrix converts a matrix into a vector, e.g.\ by stacking columns on top of each other. In the case of a symmetric matrix of size $n\times n$, only the $n\,(n+1)/2$ independent entries are stacked.}, acting on redshift indices of the symmetric tomographic matrix, and ${\tens\Sigma}_n^{-1}$ is the inverse of the covariance matrix in a given multipole bin, $\tens\Sigma_n=\langle\vec r_n^X\,(\vec r_n^X)^{\sf T}\rangle$.

The covariance (and thus also the mock realisations of the spectra) includes all possible redshift correlations and is calculated for both survey configurations described above. It is often useful to consider the $\chi^2$ in the context of the number of degrees of freedom (DOF). In this case, there are $N=12$ chosen covariance bins, which effectively act as our `data points', and $n$ redshift bins, leading to $n(n+1)/2$ independent spectra of equal and unequal redshift correlations. Thus, we can expect ${\rm DOF}_{\rm DR1} = N\times n\times(n+1)/2= 252$ for DR1 and ${\rm DOF}_{\rm DR3} = 1092$ for DR3.

\section{\label{sc:res} Results}
We now present the main results of our analysis. The aim of this section is to quantify the performance of the flat-sky approximation and Limber approximation \citep[for the latter, see also][]{Paper1,Paper2} in the context of \Euclid (both for DR1 and DR3 set-ups), assessing how both perform compared to the exact solution.

\subsection{Power spectrum recovery}

\begin{figure}[ht]
\includegraphics[scale=0.325]{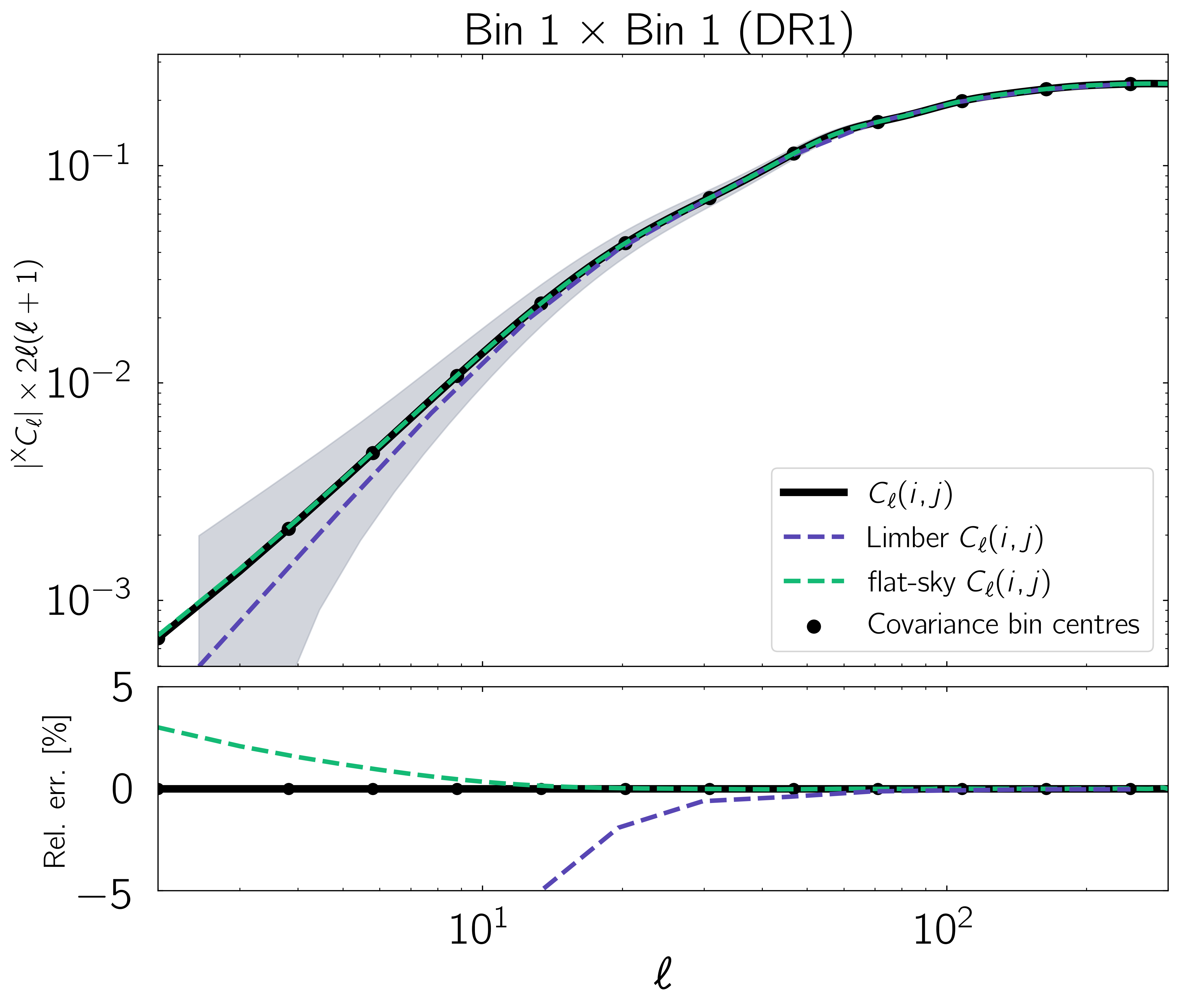}
\caption{\textit{Upper Panel:} Equal redshift angular power spectrum for the lowest redshift bin of \Euclid, using six equi-populated bins. In black solid line we show the full calculation result from \texttt{CLASS}, and in dashed purple the Limber approximation result. The green dashed line corresponds to the recalibrated version of the flat-sky approximation. \textit{Lower Panel:} Relative error (in \%) of each approximation from the full calculation result.}
\label{EucBin1x1of6}
\end{figure}

\begin{figure}[ht]
\includegraphics[scale=0.325]{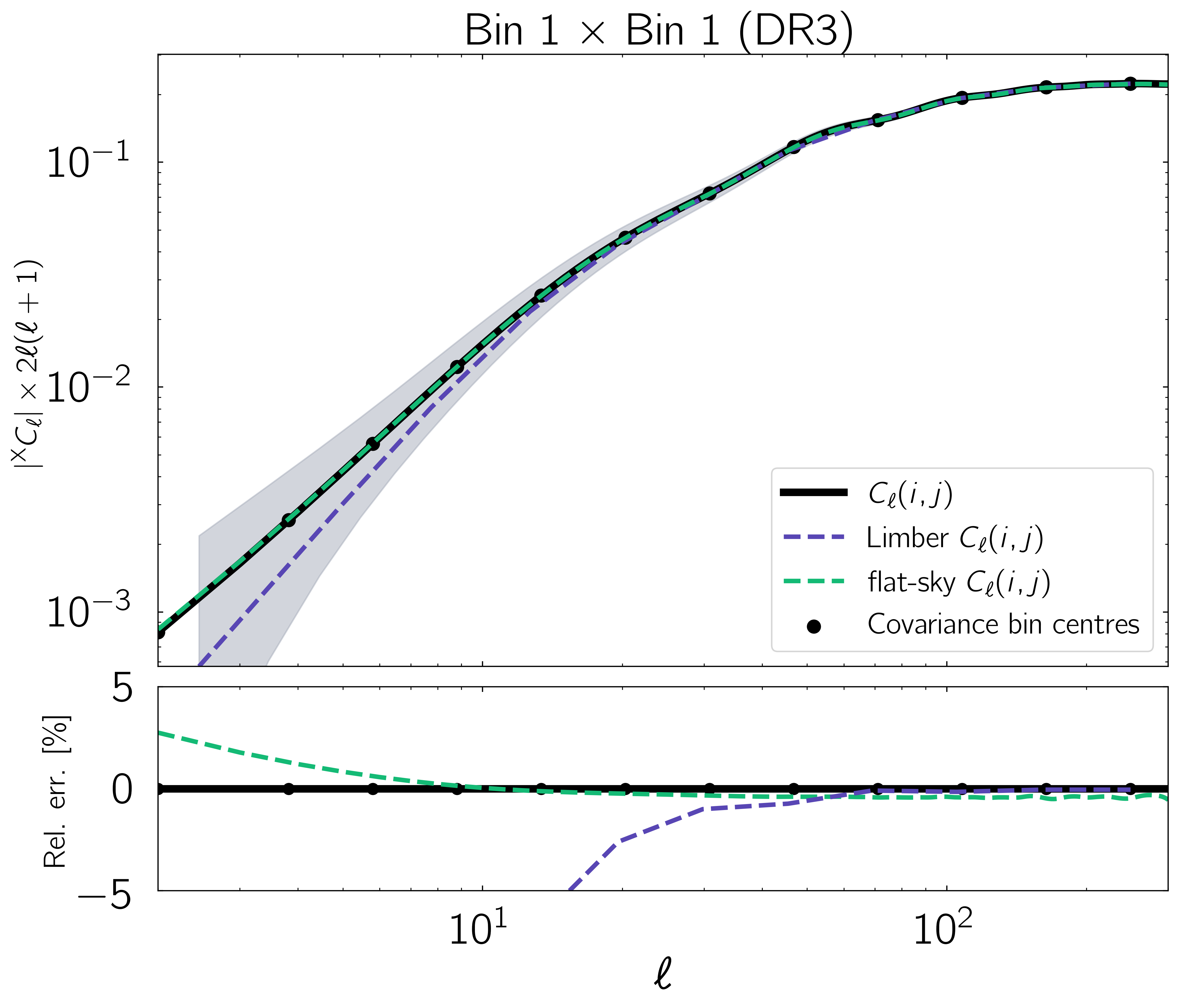}
\caption{\textit{Upper Panel:} Equal redshift angular power spectrum for the lowest redshift bin of \Euclid, using 13 equi-populated bins. In black solid line we show the full calculation result from \texttt{CLASS}, and in dashed purple the Limber approximation result. The green dashed line corresponds to the recalibrated version of the flat-sky approximation. \textit{Lower Panel:} Relative error (in \%) of each approximation from the full calculation result.}
\label{EucBin1x1of13}
\end{figure}

In \cref{EucBin1x1of6,EucBin1x1of13} we plot results for the tomographic angular power spectra for the first bin of the DR1 and DR3 configurations, respectively.  The true result taken from a full-sky calculation of \texttt{CLASS} is displayed, along with the results using the Limber and flat-sky approximations. The relative error with respect to the full calculation at each multipole is shown in the lower panels.

\begin{figure}[ht]
\includegraphics[scale=0.36]{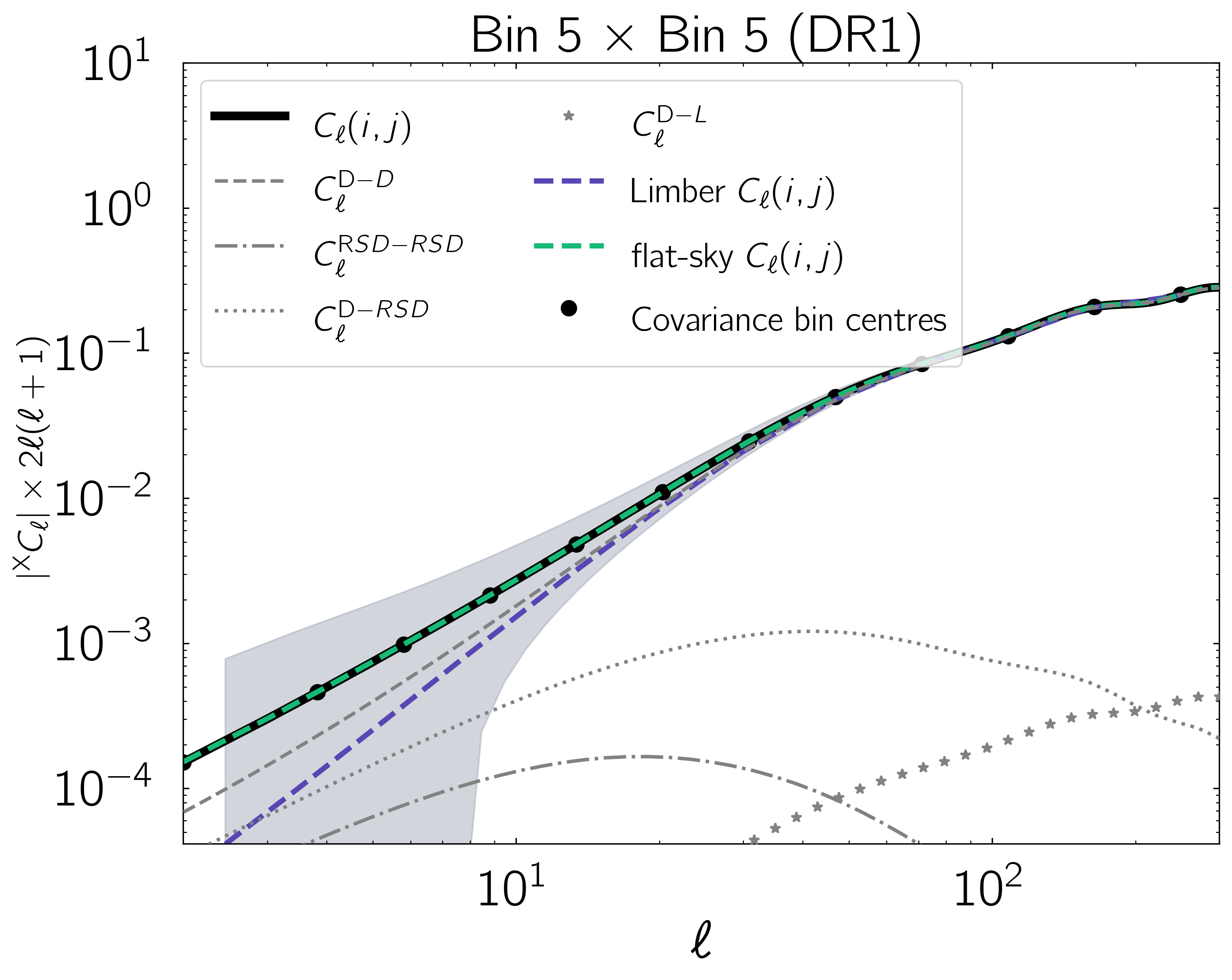}
\caption{Equal redshift angular power spectrum for the $5^{\rm th}$ redshift bin of \Euclid (DR1). In black solid line we show the full calculation result from \texttt{CLASS}, and in dashed purple the Limber approximation result. The green dashed line corresponds to the recalibrated version of the flat-sky approximation. The various grey lines represent the contributions, from observational effects, to the exact solution (calculated using \texttt{CLASS}). The L-L contribution in this correlation is too insignificant to be visible on these axes.}
\label{f:contrib}
\end{figure}

We calculate only until $\ell=300$ since, on smaller scales than this, the Limber approximation performs sufficiently well to recover the full calculation result without bias 
and it is always well within the grey shaded
$1\sigma$ error band for the survey specifications considered here. We also include an estimate of the error on the observed angular power spectrum, shown as $1\sigma$ contours in light grey. In addition, we show an example of the observational effect contributions to the angular power spectrum of DR1 in \cref{f:contrib}, for the full calculation from \texttt{CLASS}.

\begin{figure}[ht]
\includegraphics[scale=0.325]{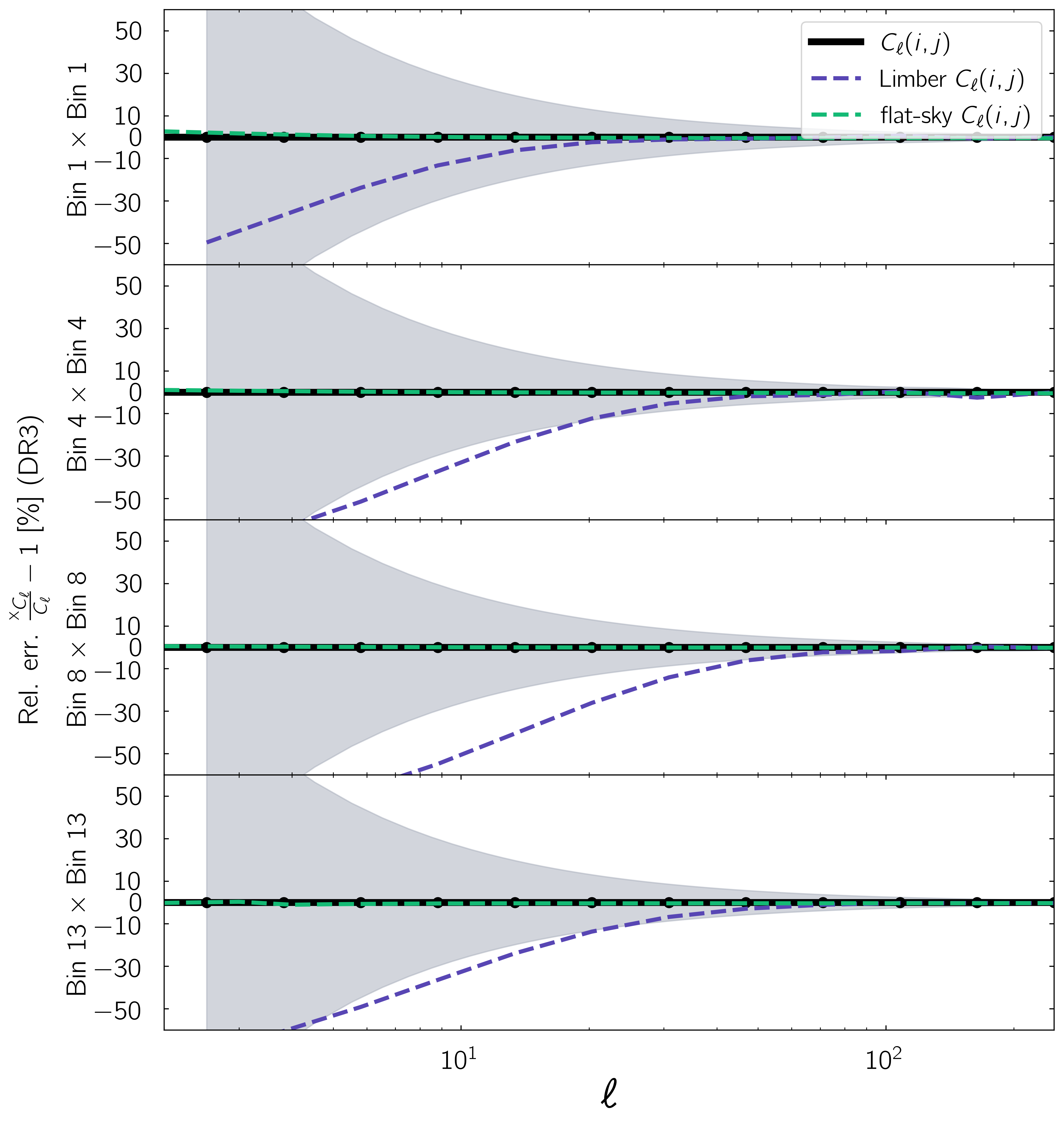}
\caption{Relative errors in the equal redshift angular power spectrum for the $1^{\rm st}$, $4^{\rm th}$, $8^{\rm th}$, and $13^{\rm th}$ redshift bins of \Euclid, using 13 equi-populated bins (DR3). In black solid line we show the full calculation result from \texttt{CLASS}, and in dashed purple the Limber approximation result. The green dashed line corresponds to the recalibrated version of the flat-sky approximation. The relative error associated with the Gaussian covariance calculated at $1\sigma$ for the given survey configuration is shown in the grey contour. Note that the $y$-axis ranges are $[-50,50]$, in contrast to \cref{EucBin1x1of6,EucBin1x1of13}.}
\label{EucBin1n3n13of13}
\end{figure}

\begin{figure}[ht]
\includegraphics[scale=0.325]{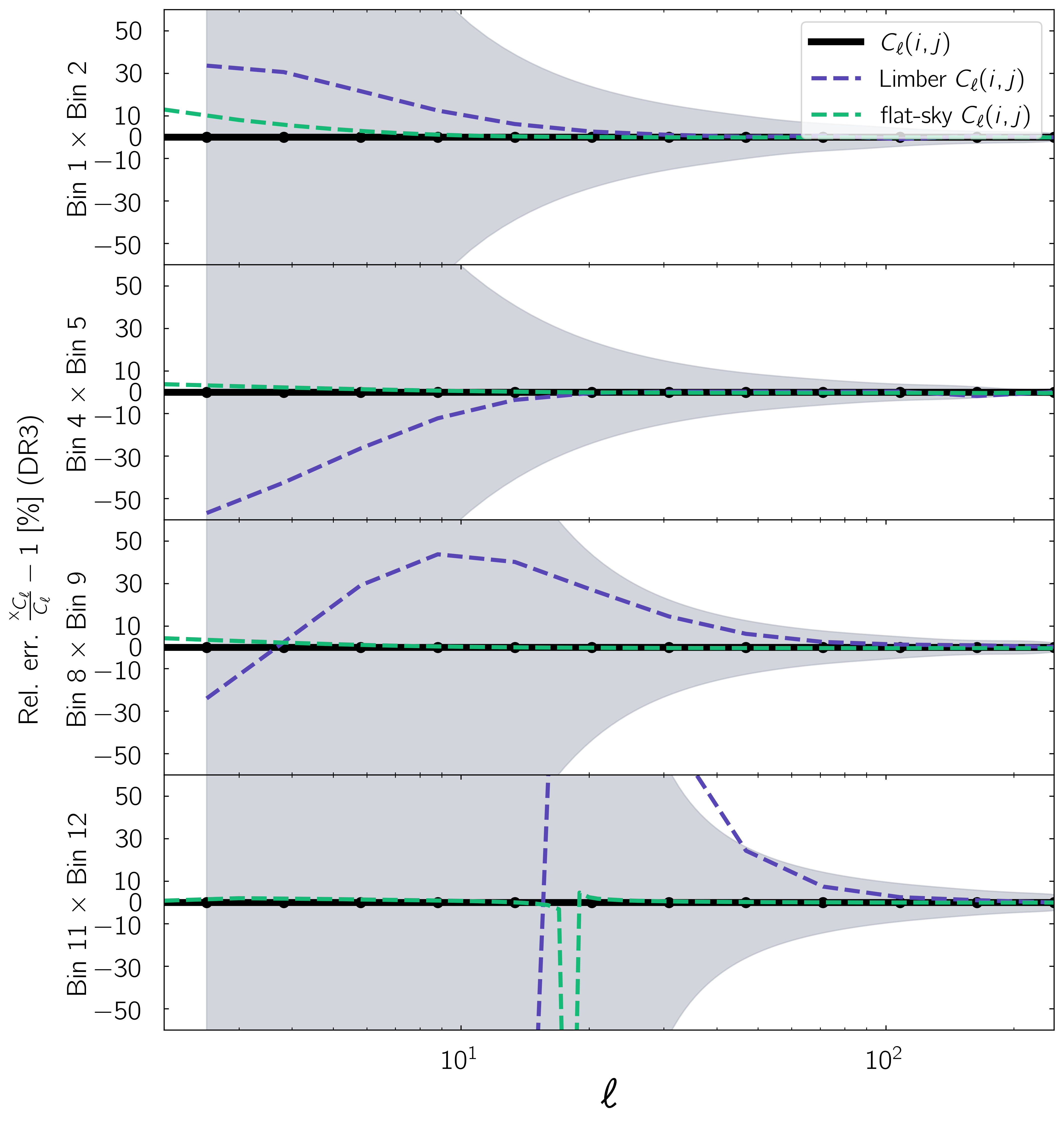}
\caption{Same as in \cref{EucBin1n3n13of13}, but now for un-equal redshift bin correlations $1\times2$, $4\times5$, $8\times9$ and $11\times12$ from DR3. {The near vertical excursions in the  $11\times 12$ correlation is due to a zero crossing of the true correlation spectrum.}}
\label{EucBin12n78n1112of13}
\end{figure}

\Cref{EucBin1n3n13of13,EucBin12n78n1112of13}, instead, show the relative errors of the approximations for various auto- and cross-bin correlations of the DR3 configuration.
In general, the {relative errors} grow, for unequal redshift correlations between increasingly disparate bins, due to the sensitivity required to detect the smaller signals. In addition, the relative errors do not vary much for equal-$z$ with increasing redshift, though the Limber approximations for the density and redshift terms become worse. This can be understood as a fixed angular scale corresponding to larger co-moving separations as redshift increases, and thus a poorer recovery of the spectrum by the Limber approximation, which completely neglects $k_\|$ --  an approximation that becomes worse when $k_\perp =\ell/r(z)$ gets smaller.\footnote{At larger scales, where na\"ive geometric intuition seems to suggest that the flat-sky approximation should also fail, it is actually impressively accurate. This has been noted before \citet{Matthewson:2020rdt}.} We use this to inform our choice of correlations shown, and in \cref{EucBin1n3n13of13}, for the case of DR3, we include correlations in redshift where Limber first becomes significantly different to the full result (Bin 4 $\times$ Bin 4), where it reaches one of the most significant departures (Bin 8 $\times$ Bin 8), as well as the first bin and last bin equal-$z$ correlations.

Overall, the results from the DR1 configuration show that the improved accuracy of the flat-sky approximation does not result in any useful improvement compared to the Limber approximation in the context of the expected observational error: i.e., it will not be possible to measure the spectra accurately enough to distinguish between the two approximation results. This is true across the entire redshift range, and for correlations between any two redshift bins, though we do not show all these spectra here, for the sake of brevity. This is expected, since the DR1 footprint will be about \(10\)--\(15\,\%\) of the full area covered by DR3, meaning that the largest scales, where the discrepancy between \cref{e:CAlijfull,e:CAlijLimber,e:flatskyfin} is more apparent, will be undersampled with respect to the final data release.

{However, in the DR3 configuration, the situation is} different. Starting as low as the $4^{\rm th}$ redshift bin (see second panel from the top in \cref{EucBin1n3n13of13}), the Limber approximation fails to recover the spectrum well enough to fall within the $1\sigma$ contour. This is also true for the equal-redshift spectra in all of the higher redshift bins, and occurs at relatively low $\ell$, where we would expect Limber to encounter difficulty in approximating the spectrum. As in the DR1 case, the Limber approximation of neighbouring bin cross correlations are not, for the most part, significantly different to the flat-sky approximation results, falling within the $1\sigma$ contour for the survey in all but one case (Bin 11 $\times$ Bin 12, which is itself merely due to a zero crossing in the full calculation correlation spectrum). 

If we consider the individual contributions to the spectra from observational effects, it may provide a hint as to why, in particular, the Limber approximation is insufficient for the full \Euclid data release.
For the DR1 case (see, e.g., \cref{f:contrib}), the largest contributions\footnote{A reminder to the reader that we use the shorthand D for density, RSD for redshift space distortions, and L for lensing. Here we are referring to cross correlations between these terms. For example, `D-D' refers to the density-density contribution to the angular power spectrum of galaxy number counts.} are D-D followed by D-L (more important at higher redshift, and correlations between large redshift separated bins). L-L also increases at higher redshift, but is still only of negligible secondary significance with respect to the former two. {As expected for a photometric survey, RSDs are not very relevant.} RSD-RSD is always smaller than D-D and for each correlation the largest contributions from D-RSD and D-L are of comparable magnitude, with D-RSD being more important at lower $\ell$ and closer $z$ correlations.

The DR3 case (not shown) is very similar. The major difference is that {due to the slimmer redshift bins, the} RSD-RSD contribution can be comparable to D-D and D-RSD in the regimes where they are important, i.e.\ at large scales in correlations of nearby bins \citep[][]{Tanidis_2019,Tanidis-TBD}.  The increased overall inaccuracy of the Limber approximation is likely caused by the failure to accurately recover the RSD terms, as a result.

\subsection{
Simulation results
}
We now assess which, if either, of the two approximations is best to model \Euclid data, using a $\chi^2$ test. This is done by computing the distance between the correct computation of the tomographic harmonic-space power spectra, as per \cref{e:CAlijfull}, and the results obtained with either approximation \cref{e:CAlijLimber,e:flatskyfin}. The advantage of this approach is that it takes into account all the possible correlations between different bins, across a range of multipoles $\ell$, allowing for the evaluation of the performance as a whole, which is not as easy through simple inspection of the individual spectra alone.

As outlined in Sect.\ \ref{sec:assess}, we calculate the $\chi^2$ distribution of each approximation, given the expected error budget of each survey configuration, arising from 1000 Gaussian realisations of measurement noise. Specifically, we adopt each simulation as a synthetic, noisy data set, and compute the $\chi^2_X$ for both approximations, as per \cref{eq:X2}. The further apart these distributions are from the one produced for the \texttt{CLASS} full-sky result, $\chi^2_{\rm true},$ the more sensitive the observations are to differences between the calculated spectra and the full calculation.

We examine the distribution of the true model to verify that the framework of our analysis is returning a reasonable fit to the data, and to give a benchmark for the expected width of the distribution in the case of realistic data. In order to judge whether the model is a good enough description of the data, we can determine the number of degrees of freedom and check whether the mean of the $\chi^2$ distribution lies close to this; i.e., $\bar\chi^2/{\rm DOF} \sim 1$. 
The overlap in distributions, or $\Delta\chi^2>0$, can be associated with a $p$-value, which acts as a frequentist measure of the probability that each distribution can achieve the $\chi^2$ of the full calculation, subject to data uncertainties. Since we run only 1000 noisy realisations, our smallest possible $p$-value is $0.001$. Our null hypothesis is that the $\Delta\chi^2$ distribution for a particular approximation is consistent with zero. That is, the mean of the $\chi^2$ distribution of the approximation is not significantly different to (in this particular case, greater than) that of the full calculation. At the $95\%$ confidence limit, the null hypothesis will be rejected for a $p$-value less than 0.05, and the distributions can be said not to be consistent at that confidence level. The $\Delta\chi^2$ distribution is particularly useful here, because each sample in the distribution corresponds to the $\Delta\chi^2$ values for corresponding noisy realisations.

In both binning configurations, the full calculation produces a distribution with mean $\bar\chi^2/{\rm DOF}\sim1$, indicating that the simulated `data' are reasonable realisations of the true result. In the six equi-populated bin configuration, the mean of the true \texttt{CLASS} distribution is at $\bar\chi^2/{\rm DOF}_{\rm DR1} \sim 0.9964$, while for the 13 bin configuration the mean of the true \texttt{CLASS} distribution is at $\bar\chi^2/{\rm DOF}_{\rm DR1} \sim 1.000$.

\begin{figure}[ht]
\includegraphics[width=\columnwidth]{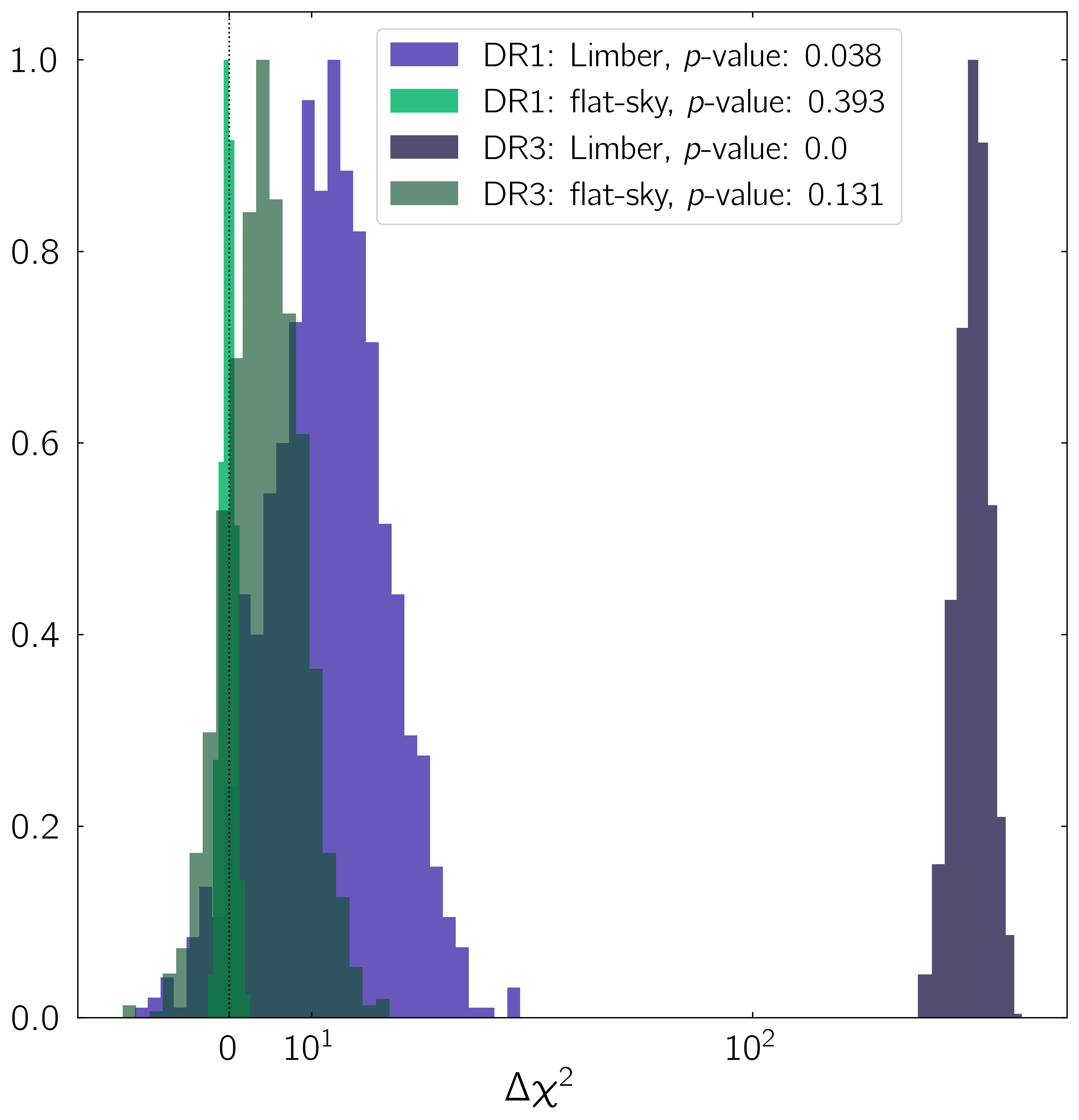}
\caption{The $\Delta\chi^2$ frequency distributions of the flat-sky (green) and Limber (purple) approximations compared to the full result, with the spread provided by the Gaussian covariance for each survey configuration. The DR1 bin configuration is shown in lighter colours, while the eventual DR3 bin result is shown in darker colours. The improvement of the flat-sky over the Limber approximation is seen in the former's distributions' closer proximity to the dotted vertical black line at $\Delta\chi^2 = 0$, most notable in the DR3 bin case. \textit{Note:} The histograms are each normalised to their own peak values and the $x$-axis is log-scaled above $\Delta\chi^2=50$.}
\label{Dchi2-FSvsFULL}
\end{figure}

In \cref{Dchi2-FSvsFULL} we show the distribution for the 1000 values of $\Delta\chi^2_X\coloneqq\chi^2_X-\chi^2_{\rm true}$. For the DR1 case, the result with the Limber approximation includes the dotted zero line, peaking around $\Delta\chi^2 \sim 12$, whilst the flat-sky approximation only incurs a $\Delta\chi^2 \sim 0.2$. The $p$-values for these cases are $p_{\rm Limber} = 0.038$ and $p_{\rm flat-sky} = 0.393$, respectively. This means that for the Limber approximation, the null hypothesis is rejected at the $95\%$ confidence level. The flat-sky approximation result is not sufficient to reject the null hypothesis that the $\Delta\chi^2$ distribution is consistent with zero.

Comparatively, for the case of the DR3 data, the distinction between the Limber and flat-sky approximations is significantly larger. In terms of the $\Delta\chi^2$, \cref{Dchi2-FSvsFULL} shows that the flat-sky approximation only reduces the overall recovery of the full calculation result by $\Delta\chi^2 \sim 5$ on average, with $p_{\rm flat-sky} = 0.131$, indicating that the null hypothesis that the distribution remains compatible with $\Delta\chi^2=0$ still cannot be rejected at $95\%$ confidence. In contrast, the Limber approximation results in a degradation with a distribution which now peaks around $\Delta\chi^2\sim 410$, with a $p$-value at the limit of our number of simulations ($p_{\rm Limber}<0.001$), rejecting the null hypothesis at $99\%$. 

\section{\label{sc:con} Discussion}
In this paper we compared the performance of the Limber and flat-sky approximations in the the determination of the angular power spectra of galaxy number counts from the photometric survey of \Euclid. We performed this study in order to assess whether the Limber approximation that enters present standard analysis packages for \Euclid, like CLOE \citep{Paper1,Paper2} performs well enough to use for \Euclid data. We showed that for the six bin analysis foreseen with DR1, the Limber approximation is sufficient to model the angular power spectrum. While leading to a slight degradation of $\chi^2$ as compared to the full or the flat-sky analysis, it is still able to recover the true result to within the $1\sigma$ error, given the survey configuration. This is no longer the case for the DR3, which will be analysed in 13 bins. These bins are sufficiently narrow to render the Limber approximation too crude, with a $p$-value below $0.1\%$. In this case the inaccuracies exceed the $1\sigma$ errors in the individual equal redshift correlations, resulting in a biased $\Delta\chi^2$ distribution, far-removed from what can reasonably be expected from the covariance in the case of the full calculation. In contrast, the flat-sky approximation is able to maintain accurate recovery of the power spectra, with a $p$-value of $13.1\%$ at worst, indicating that the overall error introduced by the approximation is small enough to avoid biasing the $\Delta\chi^2$ distribution.

An accurate approximation is only useful if it can also be used to gain computational speed, compared to the full analysis. Since this work does not promote a particular code, we did not conduct an in-depth analysis of the numerical efficiency here. Nevertheless, it is useful to briefly relay the relevant computational timings encountered, to give an idea of the current state of the problem. The 10-fold improvement in performance compared to the full calculation, as reported in \citet{Gao:2023tcd}, degrades when dealing with wide photometric windows, which require finer meshgrids to achieve accurate results. We find, for our modified version of the flat-sky code, a speed-up of a little over 2 times on average, with respect to the full calculation, in order to achieve the level of accuracy described in this work. This is still much slower than the Limber approximation, which is, on average, 200 times faster than the full calculation in this case.

{As the flat-sky calculation still involves a triple integral, a speed up along the lines of the \texttt{BLAST} code \citep{Chiarenza:2024rgk} will be required. However, further optimisations to the flat-sky code are still possible, for example, by optimising the sampling of the radial window functions~\citep{Gao:2023tcd}. As demonstrated here, the largest deviations of the Limber approximation occur for $\ell\lesssim20$ so, in practice, some combination of the two approximations could be used, depending on the relevant scales of the application. Furthermore, for the D-L and L-L contribution, the Limber approximation remains sufficiently accurate, i.e. with deviation less than a few \% of cosmic variance from the full numerical result, see~\citet{Matthewson:2020rdt}, so that it can certainly be used for well-separated bins where these terms dominate, though the total signal in these bins is small. Indeed, most of the unequal power spectra amplitudes have a contribution that is less than 30$\%$ of the corresponding equal redshift amplitude, and never exceeds 50$\%$, which it attains only for the case of neighbouring bin correlation $7 \times 8$. Finally, the $\ell$-dependence of the D-D and RSD-RSD equal redshift terms can be cast as a simple dependence in the lower boundary of the $k$-integration. Even though the spectrum calculation requires a triple integral, as explained in detail in \cite{Gao:2023tcd}, the inner $k$-integration can be pre-computed, which means that the computation time in the case of MCMC parameter estimation would be reduced by on the order of a power of $2/3$ for each point in the explored parameter space}.

{Let us finally address the relevance of this finding for the spectroscopic survey. The spectroscopic number counts are usually analysed by determining the 3D power spectrum which does not employ the Limber approximation. However, a 6\texttimes2pt analysis, including correlations of the spectroscopic number counts and the shear, is also planned. When using Limber, either one should use wide redshift bins for the spectroscopic survey also, in which case the 6\texttimes2 analysis does not add anything to the constraints (see \citealt{EP-Paganin}), or one should choose slim redshift bins. For the redshift bins considered for DR3, we confirm from the results presented in this work that the Limber approximation will be insufficient, and that the flat-sky approximation provides a promising alternative.} 

\begin{acknowledgements}
RD and WM are partially supported by the Swiss National Science Foundation. SC acknowledges support from the Italian Ministry of University and Research (\textsc{mur}), PRIN 2022 `EXSKALIBUR – Euclid-Cross-SKA: Likelihood Inference Building for Universe's Research', Grant No.\ 20222BBYB9, CUP D53D2300252 0006, from the Italian Ministry of Foreign Affairs and International
Cooperation (\textsc{maeci}), Grant No.\ ZA23GR03, and from the European Union -- Next Generation EU. IT has been supported by the Ramon y Cajal fellowship (RYC2023-045531-I) funded by the State Research Agency of the Spanish Ministerio de Ciencia, Innovaci\'on y Universidades, MICIU/AEI/10.13039/501100011033/, and Social European Funds plus (FSE+). IT also acknowledges support from the same ministry, via projects PID2019-11317GB, PID2022-141079NB, PID2022-138896NB; the European Research Executive Agency HORIZON-MSCA-2021-SE-01 Research and Innovation programme under the Marie Sk\l odowska-Curie grant agreement number 101086388 (LACEGAL) and the programme Unidad de Excelencia Mar\'{\i}a de Maeztu, project CEX2020-001058-M. 
\AckCosmoHub
\AckEC  
\end{acknowledgements}

\bibliography{Euclid}
\end{document}